\documentclass[11pt,a4paper]{article}
\usepackage[T1]{fontenc}
\usepackage{lmodern}
\usepackage{xcolor}
\usepackage[margin=26mm,headheight=14pt]{geometry}
\usepackage{amsmath,amssymb,amsthm,mathtools}
\usepackage{array,booktabs}
\usepackage{microtype}
\usepackage{enumitem}
\usepackage{fancyhdr}
\usepackage{needspace}
\usepackage[
  backend=biber,
  style=numeric,
  sorting=none,
  maxbibnames=99,
  doi=true,
  url=true,
  eprint=true
]{biblatex}

\usepackage[hidelinks]{hyperref}
\usepackage{cleveref}
\allowdisplaybreaks[1]
\newtheorem{theorem}{Theorem}[section]
\newtheorem{proposition}[theorem]{Proposition}
\newtheorem{lemma}[theorem]{Lemma}
\newtheorem{corollary}[theorem]{Corollary}
\theoremstyle{remark}

\crefname{theorem}{Theorem}{Theorems}
\crefname{lemma}{Lemma}{Lemmas}
\crefname{proposition}{Proposition}{Propositions}
\crefname{corollary}{Corollary}{Corollaries}

\usepackage{authblk}

\title{An Entanglement Monotone for GHZ Distillation}

\author[1,2]{Shashank Shekhar}
\author[2]{David Elkouss}

\affil[1]{Department of Physics, Indian Institute of Technology Bombay, Powai, Mumbai 400076, India}

\affil[2]{Networked Quantum Devices Unit, Okinawa Institute of Science and Technology Graduate University, Onna, Okinawa 904-0495, Japan}
\date{}
\begin{document}

\maketitle

\begin{abstract}
Distributing multipartite entanglement is central to many applications of quantum networks, yet determining how much can be distilled from a given quantum resource shared among the nodes remains challenging. We introduce the minimum measured correlation entanglement $U(\rho)$, a partition-based monotone that bounds Greenberger--Horne--Zeilinger (GHZ) distillation from arbitrary multipartite states under separable operations, and hence under local operations and classical communication (LOCC). It bounds the exact deterministic GHZ yield, while a finite-error estimate controls approximate preparation and bounds vanishing-error distillation rates by the regularized quantity $U_\infty(\rho)$. For networks of independent
bipartite pure states, we evaluate the bound in terms of the edge entanglement entropies. For integer Bell-pair supplies, spanning-tree packing and local fusion attain its integer part, giving the optimal
exact deterministic yield; entanglement concentration gives matching asymptotic rates for arbitrary pure edges. We also determine the optimal asymptotic GHZ rate for a family of locally flagged correlated Bell networks. Finally, a direct fidelity argument proves a strong converse at the independent pure-edge capacity: the target fidelity vanishes at every fixed rate above capacity, with explicit exponential decay for maximally entangled edges.
\end{abstract}

\section{Introduction}
\label{sec:introduction}

Quantum networks enable distribution of entanglement among distant nodes for
distributed information processing~\cite{Kimble2008,WehnerEtAl2018}. Many such applications rely on multipartite entangled states shared among multiple network nodes. The
Greenberger--Horne--Zeilinger (GHZ) state~\cite{GreenbergerEtAl1990},
for example, supports quantum conference key agreement, quantum secret
sharing, and distributed clock sensing~\cite{EppingEtAl2017,HilleryEtAl1999,KomarEtAl2014}. Generating such states requires protocols that combine the short-range entanglement supplied by the network links.

The allowed transformations depend on both the joint resource state and the allocation of its subsystems to the parties. We ask: \emph{How many GHZ states shared by all parties can be distilled from a given distributed quantum resource under local operations and classical communication (LOCC)?} Each party may process all its
registers jointly, including across copies, with adaptive operations and unrestricted classical communication.

Constructive approaches exploit network structure to distribute
GHZ and more general graph states~\cite{MeignantEtAl2019,
WallnoeferEtAl2019,Dur20162DQRintro,shekhar2026MQR}.
Multipartite distillation protocols, combinatorial methods,
and entanglement-conversion techniques provide achievable GHZ
rates and bounds on multipartite transformations~\cite{
SalekWinter2022,SalekWinter2025,VranaChristandl2019,
StreltsovEtAl2020}.
Squashed-entanglement methods give converse bounds on entanglement
and key distribution over quantum channel
networks~\cite{AzumaMizutaniLo2016}. Their multipartite extension
uses partitions of the network nodes to bound GHZ and multipartite
private-state distribution over quantum broadcast networks,
including a contribution from the initially shared
state~\cite{BaeumlAzuma2017}.
For quantum channel networks, linear-programming methods give
upper and lower bounds on entanglement and key distribution
rates, with achievable GHZ distribution rates obtained through
Steiner-tree packing~\cite{BaeumlEtAl2020}.
Entanglement-based approaches also provide general converse
bounds, including finite-copy limits on GHZ distillation from
arbitrary multipartite states~\cite{DasEtAl2021}.

Multipartite generalizations of the Rains relative entropy provide one-shot and asymptotic bounds on pure-state distillation from arbitrary inputs, including strong-converse bounds for
GHZ targets~\cite{MurrayEtAl2026}. For Bell-pair networks, the Rains, monsoon, and hurricane entanglement equal the minimum bipartition cut. The partition-based bound developed here can
be strictly tighter: for a triangle with one Bell pair per edge, it gives the optimal asymptotic GHZ rate $3/2$, whereas the minimum-cut value is $2$.

Related capacity formulas arise in conference-key generation. For independent pairwise secret-key networks, spanning-tree packing achieves the conference-key capacity under public
authenticated classical communication~\cite{TrushechkinEtAl2026}.
This result concerns classical processing of previously established keys. Closely related work determines the conference-key and GHZ distillation capacities of networks of independent GHZ resources, including nonuniform
states~\cite{Sun2026}. Its specialization to pure bipartite resources gives the same entropy-weighted GHZ capacity formula obtained here.

In this work, we introduce entanglement monotones that bound GHZ distillation from arbitrary distributed quantum resources. For each partition
$\mathcal P$ of the network nodes, we define the \textit{partition measured correlation entanglement} $E_{\mathcal P}(\rho)$ by combining relative entropy under restricted measurements~\cite{Piani2009} with a mixed convex
roof~\cite{YangEtAl2009}. Its minimum over nontrivial partitions defines the \textit{minimum measured correlation entanglement} $U(\rho)$. Both quantities are monotone under separable operations and hence under LOCC. We show that $U(\rho)$ bounds the exact deterministic GHZ yield through its normalization on GHZ states. We then establish a finite-error converse for approximate preparation, from which the regularized quantity $U_\infty(\rho)$ bounds the asymptotic distillation rate with vanishing error.

For independent pure bipartite edges, we evaluate the monotones in
terms of the edge entanglement entropies. For integer Bell-pair
supplies, spanning-tree
packing~\cite{Tutte1961,NashWilliams1961} and local
fusion~\cite{MeignantEtAl2019} attain the integer part of the
bound, giving the optimal exact deterministic GHZ yield.
For arbitrary pure edges, entanglement
concentration~\cite{BennettEtAl1996} followed by packing and
fusion attains the asymptotic bound, recovering the
entropy-weighted GHZ capacity within this monotone framework.

For locally flagged correlated Bell-network mixtures, reading the
flags and pooling Bell pairs across copies attains the bound,
giving the optimal asymptotic GHZ rate. This family includes
biseparable states with positive GHZ distillation rate,
connecting the result to multicopy activation of genuine
multipartite entanglement~\cite{YamasakiEtAl2022,
PalazuelosDeVicente2022}. Finally, a direct fidelity argument
establishes a strong converse at the independent pure-edge
capacity: at every fixed rate above it, the fidelity
with the complete GHZ target tends to zero, even under
separable operations. For maximally entangled edges, we obtain
an explicit exponential fidelity bound.

\section{Main Results}
\label{sec:operational-setting}

Let $V=\{1,\ldots,N\}$, with $N\geq2$, label the physical parties. Their joint input is a density operator $\rho$ on $\bigotimes_{v\in V}A_v$, where every local Hilbert space is finite-dimensional.  All logarithms are base two.

\subsection{Partitions and allowed operations}
\label{subsec:partitions-operations}

A partition $\mathcal P=\{B_1,\ldots,B_q\}$ divides $V$ into $q\geq2$
nonempty blocks. Write $A_{B_i}=\bigotimes_{v\in B_i}A_v$ and allow
joint operations within each block. This enlarges the operation class,
so any converse for the grouped parties also applies to the original
parties.

The parties may use local operations and classical communication
(LOCC), including local ancillas, adaptive measurements, and discarding,
without additional shared entanglement. We also consider the larger
class of separable channels. A channel separable across $\mathcal P$
admits a product-Kraus representation
\begin{equation}
    \Lambda(X)=\sum_z K_z X K_z^\dagger,
    \qquad
    K_z=\bigotimes_{i=1}^{q}K_{i,z},
    \qquad
    \sum_z K_z^\dagger K_z=I.
    \label{eq:separable-channel}
\end{equation}
Every physical LOCC channel is separable across every partition~\cite{ChitambarEtAl2014}.
Unless specified otherwise, $\mathrm{SEP}$ refers to the individual
physical parties. Conversion channels are trace preserving and discard
auxiliary output registers, leaving only the designated target registers.

\subsection{GHZ targets, yields, and distillation rates}
\label{subsec:ghz-tasks}

The all-party GHZ target~\cite{GreenbergerEtAl1990} of local dimension $d\geq2$ is
\begin{equation}
    |\Gamma_{V,d}\rangle
    =\frac{1}{\sqrt d}\sum_{x=0}^{d-1}
      \bigotimes_{v\in V}|x\rangle_v,
    \qquad
    \Gamma_{V,d}=|\Gamma_{V,d}\rangle\langle\Gamma_{V,d}|.
    \label{eq:ghz-target}
\end{equation}
In particular, $\Gamma_{V,2^k}$ is locally equivalent to $k$ qubit
GHZ states. We include $k=0$ by taking $d=1$ as the trivial target.
For a pure target $\Gamma$, we use overlap fidelity and trace distance,
\begin{equation}
    F(\sigma,\Gamma)=\operatorname{Tr}(\Gamma\sigma),
    \qquad
    T(\sigma,\omega)=\frac12\|\sigma-\omega\|_1,
    \label{eq:fidelity-trace-distance}
\end{equation}
where $\|X\|_1=\operatorname{Tr}\sqrt{X^\dagger X}$.

For $\mathcal C=\mathrm{LOCC}$ or $\mathrm{SEP}$, define the exact
deterministic yield by
\begin{equation}
    K_{\mathcal C}(\rho)
    :=\max\bigl\{k\in\mathbb N_0:
       \Lambda(\rho)=\Gamma_{V,2^k}
       \text{ for some }\Lambda\in\mathcal C\bigr\}.
    \label{eq:exact-yield}
\end{equation}
Exact conversion produces the target state on the designated registers.
A deterministic protocol is trace preserving and includes every
measurement outcome and its correction, without postselection.

Following the standard framework of asymptotic multipartite
entanglement conversion~\cite{BennettEtAl2000}, the parties receive $n$ copies of the entire input and may process them jointly. A rate $R\geq0$ is achievable
under $\mathcal C$ if there are channels $\Lambda_n\in\mathcal C$ and
integers $k_n\geq0$ such that
\begin{equation}
    T\!\left(\Lambda_n(\rho^{\otimes n}),
             \Gamma_{V,2^{k_n}}\right)\longrightarrow0,
    \qquad
    \liminf_{n\to\infty}\frac{k_n}{n}\geq R.
    \label{eq:achievable-rate}
\end{equation}
The vanishing-error rate $R_{\mathcal C}(\rho)$ is the supremum of all
achievable rates; for LOCC we also write
$R_{\mathrm{GHZ}}(\rho):=R_{\mathrm{LOCC}}(\rho)$.
The error concerns the complete GHZ output, and the rate is measured
per copy of the shared input $\rho$.

\subsection{Pairwise-independent network resources}
\label{subsec:pin-resources}

Let $G=(V,E)$ be a finite undirected multigraph without self loops. Each edge
$e=\{u,v\}$ supplies a bipartite state $\rho_e^{uv}$ on distinct
registers at its endpoints. Parallel edges are allowed and counted
separately. A pairwise-independent network (PIN) state has the form
\begin{equation}
    \rho_G=\bigotimes_{e\in E}\rho_e.
    \label{eq:pin-state}
\end{equation}
Thus independence refers to the initial edge resources. The parties
may subsequently operate jointly on all registers they hold.

A maximally entangled edge of rank $D$ supplies
\begin{equation}
    |\Phi_D\rangle
    =\frac{1}{\sqrt D}\sum_{a=0}^{D-1}|a\rangle|a\rangle,
    \qquad
    \Phi_D=|\Phi_D\rangle\langle\Phi_D|,
    \qquad
    \Psi_G=\bigotimes_{e\in E}\Phi_{D_e,e}.
    \label{eq:maximally-entangled-pin}
\end{equation}
Ranks $D_e=1$ represent product resources. If edge $e$ contains
$m_e\in\mathbb N_0$ Bell pairs, then $D_e=2^{m_e}$; write
$\Psi_{G,m}$ for this state and
$K_{\mathcal C}(G,m):=K_{\mathcal C}(\Psi_{G,m})$.

For a partition $\mathcal P$, let $E(\mathcal P)$ be the multiset of
edges joining different blocks. Given nonnegative edge weights
$c=(c_e)_{e\in E}$, define
\begin{equation}
    \tau(G,c)
    :=\min_{\mathcal P:\,|\mathcal P|\geq2}
      \frac{\displaystyle\sum_{e\in E(\mathcal P)}c_e}
           {|\mathcal P|-1}.
    \label{eq:partition-strength}
\end{equation}
The relevant weights are $c_e=\log D_e$ for maximally entangled edges. For pure edge states
$\rho_e=\phi_e:=|\phi_e\rangle\langle\phi_e|$, we choose
$c_e=s_e$, where $s_e$ is the entropy of
entanglement~\cite{BennettEtAl1996},
\begin{equation}
    s_e=S\!\left(\operatorname{Tr}_v\phi_e\right),
    \qquad
    S(\omega)=-\operatorname{Tr}(\omega\log\omega),
    \qquad e=\{u,v\}.
    \label{eq:edge-entanglement-entropy}
\end{equation}
For a pure edge, either endpoint gives the same entropy.

\subsection{Main results}
\label{subsec:main-results}

The monotone gives the converses below. Spanning-tree
packing~\cite{Tutte1961,NashWilliams1961} and local
fusion~\cite{MeignantEtAl2019} attain the Bell-network yield;
entanglement concentration extends rate achievability to arbitrary
pure edges.

\begin{theorem}[General GHZ bounds and network applications]
\label{thm:main-network-results}
Let $U$ denote the minimum measured correlation entanglement introduced in Section~\ref{sec:partition-monotone}, and let $U_\infty$ denote its regularization.
The following statements hold.
\begin{enumerate}
    \item \emph{General inputs.}
    For every joint input state $\rho$,
    \begin{equation}
        K_{\mathrm{LOCC}}(\rho)
        \leq K_{\mathrm{SEP}}(\rho)
        \leq\lfloor U(\rho)\rfloor.
        \label{eq:main-general-exact-converse}
    \end{equation}
    Its vanishing-error distillation rates satisfy
    \begin{equation}
        R_{\mathrm{LOCC}}(\rho)
        \leq R_{\mathrm{SEP}}(\rho)
        \leq U_\infty(\rho).
        \label{eq:main-general-rate-converse}
    \end{equation}
    Moreover, $U(\rho)=0$ if and only if $\rho$ is separable across
    at least one bipartition of the physical parties, as established
    in Proposition~\ref{prop:faithfulness}.

    \item \emph{Bell-pair networks.}
    Let $G=(V,E)$ be a finite multigraph with no self loops and with integer
    Bell-pair supplies $m=(m_e)_{e\in E}$. Then
    \begin{equation}
        K_{\mathrm{LOCC}}(G,m)
        =K_{\mathrm{SEP}}(G,m)
        =\lfloor\tau(G,m)\rfloor.
        \label{eq:main-exact-bell-yield}
    \end{equation}

    \item \emph{Independent pure-edge networks.}
    Let $G=(V,E)$ be a finite multigraph with no self loops, and suppose
    $\rho_G=\bigotimes_{e\in E}\phi_e$ has independent pure
    bipartite edges with entanglement entropies $s_e$.
    For every positive integer $n$,
    \begin{equation}
        K_{\mathrm{LOCC}}(\rho_G^{\otimes n})
        \leq K_{\mathrm{SEP}}(\rho_G^{\otimes n})
        \leq\lfloor n\tau(G,s)\rfloor.
        \label{eq:main-exact-pure-converse}
    \end{equation}
    Its vanishing-error rates satisfy
    \begin{equation}
        R_{\mathrm{LOCC}}(\rho_G)
        =R_{\mathrm{SEP}}(\rho_G)
        =\tau(G,s).
        \label{eq:main-pure-capacity}
    \end{equation}
\end{enumerate}
\end{theorem}

The Bell-network yield is exact. For arbitrary pure edges,
\eqref{eq:main-exact-pure-converse} bounds exact conversion, while
the matching rate in \eqref{eq:main-pure-capacity} allows an error
tending to zero.

\begin{theorem}[Fidelity bound and strong converse]
\label{thm:main-strong-converse}
Let $\Psi_G$ have maximally entangled edges of ranks $D_e$ and set
$c_e=\log D_e$. Every trace-preserving separable channel satisfies
\begin{equation}
    F\!\left(\Lambda(\Psi_G),\Gamma_{V,d}\right)
    \leq\min\left\{1,\frac{2^{\tau(G,c)}}{d}\right\}.
    \label{eq:main-fidelity-bound}
\end{equation}
Consequently, for $n$ copies and a target of $k_n$ qubit GHZ states,
\begin{equation}
    F_n:=F\!\left(\Lambda_n(\Psi_G^{\otimes n}),
                  \Gamma_{V,2^{k_n}}\right)
    \leq\min\{1,2^{n\tau(G,c)-k_n}\}.
    \label{eq:main-exponential-converse}
\end{equation}
In particular, $k_n/n\geq\tau(G,c)+\delta$ for a fixed $\delta>0$
implies $F_n\leq2^{-n\delta}$.

For arbitrary independent pure edges, the same strong-converse
threshold holds: if $k_n/n\geq\tau(G,s)+\delta$ for all sufficiently
large $n$ and some $\delta>0$, then
\begin{equation}
    F\!\left(\Lambda_n(\rho_G^{\otimes n}),
             \Gamma_{V,2^{k_n}}\right)\longrightarrow0
    \label{eq:main-pure-strong-converse}
\end{equation}
for every sequence of trace-preserving separable channels $\Lambda_n$.
\end{theorem}

The explicit exponential bound in \eqref{eq:main-exponential-converse}
applies to maximally entangled edges.

\section{Measured Correlation Entanglement}
\label{sec:partition-monotone}

We construct monotones indexed by partitions of the physical parties
from relative entropy under restricted measurements~\cite{Piani2009}
and a mixed convex roof~\cite{YangEtAl2009}, normalized to assign one
unit to a qubit GHZ state.

\subsection{Definitions and properties}
\label{subsec:monotone-construction}

Fix $\mathcal P=\{B_1,\ldots,B_q\}$, with $q\geq2$. For probability
distributions $p$ and $r$, write
\begin{equation}
    D(p\|r)=\sum_y p_y\log\frac{p_y}{r_y},
    \qquad
    H(p)=-\sum_y p_y\log p_y.
    \label{eq:classical-relative-entropy}
\end{equation}
Terms with $p_y=0$ vanish, while $p_y>0=r_y$ gives $D(p\|r)=+\infty$.
We use finite POVMs and finite ensembles throughout.

A POVM $M=\{M_y\}_y$ is separable across $\mathcal P$ if
$\sum_y M_y=I$ and each effect is a sum of positive product operators,
\begin{equation}
    M_y=\sum_\beta\bigotimes_{i=1}^{q}M^{(i)}_{y,\beta},
    \qquad M^{(i)}_{y,\beta}\geq0.
    \label{eq:separable-povm}
\end{equation}
Denote this measurement class by $\mathrm{SEP}(\mathcal P)$. The
separably measured relative entropy is
\begin{equation}
    D_{\mathrm{SEP}(\mathcal P)}(\sigma\|\omega)
    :=\sup_{M\in\mathrm{SEP}(\mathcal P)}
      D\!\left(
        \{\operatorname{Tr}(M_y\sigma)\}_y
        \middle\|
        \{\operatorname{Tr}(M_y\omega)\}_y
      \right).
    \label{eq:sep-measured-relative-entropy}
\end{equation}
Minimizing over product reference states gives
\begin{equation}
    J_{\mathcal P}(\sigma)
    :=\inf_{\omega_1,\ldots,\omega_q}
      D_{\mathrm{SEP}(\mathcal P)}
      \!\left(\sigma\middle\|\bigotimes_{i=1}^{q}\omega_i\right),
    \label{eq:product-reference-functional}
\end{equation}
where $\omega_i$ is a normalized state on $A_{B_i}$. This quantity
vanishes on product states but can count classical correlations in
separable mixtures. We therefore define the \emph{partition measured correlation
entanglement} associated with $\mathcal P$ by
\begin{equation}
    E_{\mathcal P}(\rho)
    :=\frac{1}{q-1}
      \inf_{\rho=\sum_\alpha p_\alpha\sigma_\alpha}
      \sum_\alpha p_\alpha J_{\mathcal P}(\sigma_\alpha).
    \label{eq:partition-monotone}
\end{equation}
The components in the infimum may be mixed. Combining approximately
minimizing ensembles shows that $E_{\mathcal P}$ is convex, while a
product-state decomposition gives
$E_{\mathcal P}(\rho)=0$ whenever $\rho$ is separable across
$\mathcal P$.

If the total input dimension is $L$, the uniform product reference
$I/L$ bounds every outcome likelihood ratio by $L$, since
$\sigma\leq I$. Thus the quantities are finite:
\begin{equation}
    0\leq E_{\mathcal P}(\rho)
    \leq\frac{J_{\mathcal P}(\rho)}{q-1}
    \leq\frac{\log L}{q-1}.
    \label{eq:monotone-finiteness}
\end{equation}
For a pure state $\psi=|\psi\rangle\langle\psi|$, every component with
positive weight in a convex decomposition equals $\psi$, so
\begin{equation}
    E_{\mathcal P}(\psi)=\frac{J_{\mathcal P}(\psi)}{q-1}.
    \label{eq:pure-state-reduction}
\end{equation}

Minimizing over all nontrivial partitions defines the
\emph{minimum measured correlation entanglement},
\begin{equation}
    U(\rho):=\min_{\mathcal P:\,|\mathcal P|\geq2}
             E_{\mathcal P}(\rho).
    \label{eq:partition-minimum}
\end{equation}
The normalization in \eqref{eq:partition-monotone} gives every partition
the same GHZ value, so the minimum selects the strongest
exact-conversion bound. We convexify at each fixed partition before
taking this minimum; $U$ need not be convex.

\subsection{Monotonicity, local invariance, and faithfulness}
\label{subsec:monotonicity-invariance}

We prove monotonicity on average over recorded outcomes, the standard
criterion for selective entanglement monotonicity~\cite{Vidal2000}.
First, we record every product Kraus outcome in
\eqref{eq:separable-channel}. Zero-probability branches contribute zero
and are omitted from conditional-state formulas.

\begin{lemma}[Refined-instrument inequality]
\label{lem:refined-instrument}
For a complete product-Kraus instrument across $\mathcal P$, let
\begin{equation}
    p_z=\operatorname{Tr}(K_z\rho K_z^\dagger),
    \qquad
    \rho_z=\frac{K_z\rho K_z^\dagger}{p_z}.
    \label{eq:refined-branch-states}
\end{equation}
Then
\begin{equation}
    J_{\mathcal P}(\rho)
    \geq\sum_z p_zJ_{\mathcal P}(\rho_z).
    \label{eq:refined-j-monotonicity}
\end{equation}
\end{lemma}

\begin{proof}
Fix a product reference $\Omega=\bigotimes_i\omega_i$ and apply the
same instrument to it. Write
\begin{equation}
    r_z=\operatorname{Tr}(K_z\Omega K_z^\dagger),
    \qquad
    \Omega_z=\frac{K_z\Omega K_z^\dagger}{r_z}
    \quad(r_z>0).
    \label{eq:reference-branch-states}
\end{equation}
Each $\Omega_z$ is product across $\mathcal P$. If $p_z>0=r_z$ for
some branch, the separable POVM $\{K_z^\dagger K_z\}_z$ gives infinite
relative entropy, making the bound for this reference immediate.
Otherwise $r_z>0$ whenever $p_z>0$.

Fix $\delta>0$. On each such branch choose a separable POVM
$\{M_{y|z}\}_y$ whose outcome distributions $P(\cdot|z)$ and
$Q(\cdot|z)$ on $\rho_z$ and $\Omega_z$ satisfy
\begin{equation}
    D\!\left(P(\cdot|z)\|Q(\cdot|z)\right)
    \geq J_{\mathcal P}(\rho_z)-\delta.
    \label{eq:conditional-measurement-choice}
\end{equation}
This is possible because $\Omega_z$ is an admissible product reference
in \eqref{eq:product-reference-functional}. Choose any separable POVM
on the remaining branches. The input effects
\begin{equation}
    L_{z,y}=K_z^\dagger M_{y|z}K_z
    \label{eq:pulled-back-povm}
\end{equation}
are separable because conjugation by a product operator preserves
positive product decompositions. Since
$\sum_{z,y}L_{z,y}=\sum_zK_z^\dagger K_z=I$, they form a POVM with
joint distributions $p_zP(y|z)$ and $r_zQ(y|z)$. The classical
relative-entropy chain rule gives
\begin{align}
    D_{\mathrm{SEP}(\mathcal P)}(\rho\|\Omega)
    &\geq D\!\left(
        \{p_zP(y|z)\}_{z,y}
        \middle\|
        \{r_zQ(y|z)\}_{z,y}\right) \notag\\
    &=D(p\|r)+\sum_zp_z
      D\!\left(P(\cdot|z)\|Q(\cdot|z)\right) \notag\\
    &\geq\sum_zp_zJ_{\mathcal P}(\rho_z)-\delta.
    \label{eq:instrument-chain-rule}
\end{align}
The last step uses $D(p\|r)\geq0$ and $\sum_zp_z=1$.
Letting $\delta\downarrow0$ and taking the infimum over product
references proves \eqref{eq:refined-j-monotonicity}.
\end{proof}

\begin{proposition}[Selective monotonicity]
\label{prop:selective-monotonicity}
For every separable instrument across $\mathcal P$, with outcome
probabilities $p_z$ and normalized outputs $\rho_z$,
\begin{equation}
    E_{\mathcal P}(\rho)
    \geq\sum_zp_zE_{\mathcal P}(\rho_z).
    \label{eq:selective-monotonicity}
\end{equation}
In particular, $E_{\mathcal P}(\Lambda(\rho))\leq
E_{\mathcal P}(\rho)$ for every trace-preserving separable channel
across $\mathcal P$.
\end{proposition}

\begin{proof}
First refine the instrument so each outcome has one product Kraus
operator. For an arbitrary ensemble
$\rho=\sum_\alpha w_\alpha\sigma_\alpha$, let $p(z|\alpha)$ and
$\sigma_{\alpha,z}$ be the branch probabilities and normalized outputs
on component $\sigma_\alpha$. Then
\begin{equation}
    p_z=\sum_\alpha w_\alpha p(z|\alpha),
    \qquad
    \rho_z=\sum_\alpha
      \frac{w_\alpha p(z|\alpha)}{p_z}\,
      \sigma_{\alpha,z}.
    \label{eq:conditional-ensemble}
\end{equation}
Convexity, \eqref{eq:monotone-finiteness}, and
Lemma~\ref{lem:refined-instrument} imply
\begin{align}
    \sum_zp_zE_{\mathcal P}(\rho_z)
    &\leq\sum_{\alpha,z}w_\alpha p(z|\alpha)
          E_{\mathcal P}(\sigma_{\alpha,z}) \notag\\
    &\leq\frac1{q-1}\sum_\alpha w_\alpha
          \sum_zp(z|\alpha)J_{\mathcal P}(\sigma_{\alpha,z})
          \notag\\
    &\leq\frac1{q-1}\sum_\alpha w_\alpha
          J_{\mathcal P}(\sigma_\alpha).
    \label{eq:convexification-monotonicity-proof}
\end{align}
Taking the infimum over input ensembles proves the refined statement.
Coarse-grained outputs are mixtures of refined outputs, so convexity
extends the inequality to arbitrary separable instruments, including
discarding all outcomes.
\end{proof}

Appending a fixed local state is reversed by discarding it. A local
isometry is reversed on its range by its inverse, extended to a channel
by preparing a fixed state on the orthogonal complement. Applying
monotonicity in both directions therefore gives, for block-local
isometries $W_i$ and fixed local states $\tau_i$,
\begin{equation}
    E_{\mathcal P}(W\rho W^\dagger)=E_{\mathcal P}(\rho),
    \qquad
    E_{\mathcal P}\!\left(\rho\otimes\bigotimes_i\tau_i\right)
    =E_{\mathcal P}(\rho),
    \qquad W=\bigotimes_iW_i.
    \label{eq:local-invariance}
\end{equation}

For an instrument separable across the physical parties,
\eqref{eq:selective-monotonicity} holds for every partition. Since
$E_{\mathcal P}(\rho_z)\geq U(\rho_z)$, minimizing the input side gives
\begin{equation}
    U(\rho)\geq\sum_zp_zU(\rho_z).
    \label{eq:u-selective-monotonicity}
\end{equation}
Thus $U$ is selectively monotone under physical separable operations
and, in particular, under LOCC. It is also invariant under physical
local isometries and the addition of fixed product ancillas.

\begin{proposition}[Faithfulness and zero set]
\label{prop:faithfulness}
For a partition $\mathcal P$, let $\mathcal S_{\mathcal P}$ denote
the set of states fully separable across its blocks. Then
\begin{equation}
    E_{\mathcal P}(\rho)=0
    \quad\Longleftrightarrow\quad
    \rho\in\mathcal S_{\mathcal P}.
    \label{eq:partition-faithfulness}
\end{equation}
Consequently, $U(\rho)=0$ if and only if $\rho$ is separable across
at least one fixed bipartition of the physical parties.
\end{proposition}

\begin{proof}
Fix $\mathcal P=\{B_1,\ldots,B_q\}$ and define the auxiliary quantity
\[
    C_{\mathcal P}(\rho)
    :=\inf_{\omega\in\mathcal S_{\mathcal P}}
      D_{\mathrm{SEP}(\mathcal P)}(\rho\|\omega).
\]
Joint convexity of classical relative entropy is preserved by the
measurement supremum in \eqref{eq:sep-measured-relative-entropy}.
Minimizing over the convex set $\mathcal S_{\mathcal P}$ therefore
makes $C_{\mathcal P}$ convex. Since product reference states belong
to $\mathcal S_{\mathcal P}$, $J_{\mathcal P}\geq C_{\mathcal P}$, and
every ensemble decomposition $\rho=\sum_\alpha p_\alpha\sigma_\alpha$
satisfies
\[
    \sum_\alpha p_\alpha J_{\mathcal P}(\sigma_\alpha)
    \geq\sum_\alpha p_\alpha C_{\mathcal P}(\sigma_\alpha)
    \geq C_{\mathcal P}(\rho).
\]
Taking the ensemble infimum gives
$(q-1)E_{\mathcal P}(\rho)\geq C_{\mathcal P}(\rho)$.

Suppose $\rho\notin\mathcal S_{\mathcal P}$. Choose a finite
informationally complete POVM on each block and take their product
$M=\{M_y\}_y$. Its outcome map
$\mathcal M(\sigma):=(\operatorname{Tr}(M_y\sigma))_y$
is injective. The set $\mathcal S_{\mathcal P}$ is compact in finite
dimensions, so
\[
    \delta:=\min_{\omega\in\mathcal S_{\mathcal P}}
      \|\mathcal M(\rho)-\mathcal M(\omega)\|_1>0.
\]
Here $\|\cdot\|_1$ is the ordinary $\ell_1$ norm of the outcome
vectors. The product POVM is separable across $\mathcal P$;
therefore Pinsker's inequality, with base-two logarithms, implies
\[
    C_{\mathcal P}(\rho)
    \geq\inf_{\omega\in\mathcal S_{\mathcal P}}
      D\bigl(\mathcal M(\rho)\|\mathcal M(\omega)\bigr)
    \geq\frac{\delta^2}{2\ln 2}>0.
\]
Thus $E_{\mathcal P}>0$ outside $\mathcal S_{\mathcal P}$; product-state
decompositions give zero on that set, as noted after
\eqref{eq:partition-monotone}.

The minimum defining $U$ runs over finitely many partitions, so it
vanishes exactly when $\rho$ is separable across some nontrivial
partition. Such a partition can be coarsened to a bipartition, and
every bipartition is included in the minimum.
\end{proof}

Biseparable mixtures may use different cuts and need not lie in the
zero set of $U$. Section~\ref{subsec:flagged-network-activation} gives
an example with $U>0$.

\subsection{GHZ normalization and the exact converse}
\label{subsec:ghz-normalization-exact-converse}

The following lower bound from local measurement statistics will also
be used to evaluate network states.

\begin{lemma}[Classical correlation bound]
\label{lem:classical-correlation-bound}
Measure the joint system of each block in a fixed orthonormal basis. Let
$p$ be the joint outcome distribution on $\sigma$, with marginals
$p_i$. Then
\begin{equation}
    J_{\mathcal P}(\sigma)
    \geq\sum_{i=1}^{q}H(p_i)-H(p).
    \label{eq:classical-correlation-bound}
\end{equation}
\end{lemma}

\begin{proof}
This product POVM is separable, and a product reference
$\bigotimes_i\omega_i$ induces a product distribution $\prod_i r_i$.
Expanding the logarithm and summing over the other coordinates gives
\begin{align}
    D\!\left(p\middle\|\prod_i r_i\right)
    &=-H(p)-\sum_i\sum_{x_i}p_i(x_i)\log r_i(x_i)
      \notag\\
    &=\sum_iH(p_i)-H(p)+\sum_iD(p_i\|r_i) \notag\\
    &\geq\sum_iH(p_i)-H(p).
    \label{eq:classical-correlation-identity}
\end{align}
The bound holds for every product reference under the support
conventions in \eqref{eq:classical-relative-entropy}. Taking the
measurement supremum and reference infimum proves the claim.
\end{proof}

\begin{proposition}[GHZ normalization]
\label{prop:ghz-normalization}
For every partition $\mathcal P$ and every integer $d\geq2$,
\begin{equation}
    E_{\mathcal P}(\Gamma_{V,d})
    =U(\Gamma_{V,d})=\log d.
    \label{eq:ghz-normalization}
\end{equation}
\end{proposition}

\begin{proof}
First consider $q$ parties, each with one $d$-dimensional
register, and their GHZ state $\Gamma_{q,d}$.

\emph{Lower bound.}
Computational-basis measurement produces a uniform common symbol.
Thus $H(p_i)=H(p)=\log d$, and
Lemma~\ref{lem:classical-correlation-bound} gives
\begin{equation}
    J_{\mathcal P}(\Gamma_{q,d})\geq(q-1)\log d.
    \label{eq:ghz-j-lower-bound}
\end{equation}

\emph{Upper bound.}
For normalized local vectors $u_i$, H\"older's inequality gives
\[
    \left|\sum_x\prod_i\langle u_i|x\rangle\right|
    \leq\prod_i
      \left(\sum_x|\langle u_i|x\rangle|^q\right)^{1/q}
    \leq1,
\]
where the last inequality uses $q\geq2$ and
$\|u_i\|_q\leq\|u_i\|_2=1$ for the coefficient vectors. Hence
\begin{equation}
    \left|\left\langle\bigotimes_i u_i
      \middle|\Gamma_{q,d}\right\rangle\right|^2\leq\frac1d.
    \label{eq:ghz-product-overlap}
\end{equation}
Spectrally decomposing the local factors of a separable positive
effect gives $M=\sum_\beta c_\beta|u_\beta\rangle\langle u_\beta|$,
where $c_\beta\geq0$ and each $u_\beta$ is a normalized product vector.
Therefore
\begin{equation}
    \operatorname{Tr}(M\Gamma_{q,d}) = \sum_{\beta}c_{\beta}|\langle u_{\beta}|\Gamma_{q,d}\rangle|^2
    \leq\frac{\operatorname{Tr}M}{d}
    =d^{q-1}\operatorname{Tr}\!\left(M\frac{I}{d^q}\right).
    \label{eq:ghz-separable-effect-bound}
\end{equation}
For a separable POVM $\{M_y\}_y$, let
\(
    p_y:=\operatorname{Tr}(M_y\Gamma_{q,d}),
    \quad
    r_y:=\operatorname{Tr}(M_y I/d^q)
\)
be its outcome distributions on the GHZ state and product reference.
After omitting zero effects, $r_y>0$, and the preceding bound gives
\(
    \frac{p_y}{r_y}
    \leq
    d^{q-1}.
\)
Consequently, using $\sum_y p_y=1$,
\[
    D(p\|r)
    =\sum_y p_y\log\frac{p_y}{r_y}
    \leq\sum_y p_y\log d^{q-1}
    =(q-1)\log d.
\]
Taking the supremum over separable POVMs and choosing the product
reference $I/d^q$ yields
\begin{equation}
    J_{\mathcal P}(\Gamma_{q,d})
    \leq D_{\mathrm{SEP}(\mathcal P)}
       (\Gamma_{q,d}\|I/d^q)
    \leq(q-1)\log d.
    \label{eq:ghz-j-upper-bound}
\end{equation}

Combining this with \eqref{eq:ghz-j-lower-bound} and the pure-state
identity \eqref{eq:pure-state-reduction} gives
$E_{\mathcal P}(\Gamma_{q,d})=\log d$.

For the physical parties, the isometry
$|x\rangle\mapsto|x\rangle^{\otimes|B_i|}$ at each block maps
$\Gamma_{q,d}$ to $\Gamma_{V,d}$. Block-local isometry invariance
therefore gives $E_{\mathcal P}(\Gamma_{V,d})=\log d$.
Taking the partition minimum gives $U(\Gamma_{V,d})=\log d$.
\end{proof}

\begin{corollary}[Exact deterministic GHZ converse]
\label{cor:exact-ghz-converse}
Let $\rho$ be an arbitrary joint state. If a trace-preserving channel
separable across the physical parties satisfies
$\Lambda(\rho)=\Gamma_{V,d}$, then
\begin{equation}
    \log d\leq U(\rho).
    \label{eq:exact-ghz-converse}
\end{equation}
In particular,
\begin{equation}
    K_{\mathrm{LOCC}}(\rho)
    \leq K_{\mathrm{SEP}}(\rho)
    \leq\lfloor U(\rho)\rfloor.
    \label{eq:exact-yield-monotone-bound}
\end{equation}
\end{corollary}

\begin{proof}
For every partition, monotonicity and GHZ normalization give
\[
    E_{\mathcal P}(\rho)
    \geq E_{\mathcal P}(\Lambda(\rho))
    =\log d.
\]
Minimizing over partitions proves \eqref{eq:exact-ghz-converse}.
Taking $d=2^k$ and using that $k$ is an integer yields
\eqref{eq:exact-yield-monotone-bound}; the case $k=0$ is trivial.
\end{proof}

This converse requires neither a network decomposition of the input
nor an approximation argument. The following sections evaluate its
input side for independent bipartite network resources.

\section{Maximally Entangled PINs}
\label{sec:maximally-entangled-pins}

Consider the maximally entangled PIN state $\Psi_G$ defined in
\eqref{eq:maximally-entangled-pin}, with $c_e=\log D_e$. We evaluate
$E_{\mathcal P}(\Psi_G)$ for every partition to obtain an exact-conversion
bound. For integer Bell-pair supplies, spanning-tree packing and local
fusion attain its integer part.

\subsection{Graph-contraction estimate}
\label{subsec:graph-contraction}

Each edge carries a label shared by its two endpoints. The following
estimate is a finite-sum consequence of Finner's generalized
H\"older inequality~\cite{Finner1992}, with quantum counterparts for
independent bipartite sources established in Ref.~\cite{RenouEtAl2019}.
We give an elementary proof by repeated Cauchy--Schwarz.

\begin{lemma}[Graph-contraction estimate]
\label{lem:graph-contraction}
Let $H=(W,F)$ be a finite multigraph with no self loops. Assign a finite label
$a_e$ to each edge $e\in F$, and let $f_w$ be a complex array indexed
by the labels of the edges incident on $w$. Then
\begin{equation}
    \left|
      \sum_{(a_e)_{e\in F}}
      \prod_{w\in W}f_w\bigl((a_e)_{e\ni w}\bigr)
    \right|
    \leq\prod_{w\in W}\|f_w\|_2,
    \label{eq:graph-contraction-bound}
\end{equation}
where
\[
    \|f_w\|_2^2
    :=\sum_{(a_e)_{e\ni w}}
      \left|f_w\bigl((a_e)_{e\ni w}\bigr)\right|^2.
\]
An isolated vertex contributes a scalar and its absolute value.
\end{lemma}

\begin{proof}
Successively merge vertices, keeping an array indexed by the edges
leaving each group. For two groups, collect all labels of edges joining
them into $a$ and their remaining labels into $x$ and $y$. Contracting
the shared labels gives
\[
    C_{x,y}=\sum_a A_{x,a}B_{a,y}.
\]
The remaining index sets $x$ and $y$ are disjoint, because each edge
has only two endpoints. Cauchy--Schwarz gives
\begin{align}
    \|C\|_2^2
    &=\sum_{x,y}\left|\sum_a A_{x,a}B_{a,y}\right|^2 \notag\\
    &\leq\sum_{x,y}
      \left(\sum_a|A_{x,a}|^2\right)
      \left(\sum_a|B_{a,y}|^2\right) \notag\\
    &=\|A\|_2^2\|B\|_2^2.
    \label{eq:contraction-cauchy-schwarz}
\end{align}
If the groups share no edge, the merge is an outer product and the
norms multiply exactly. Contracting all shared labels, including
parallel edges, leaves precisely the labels of edges leaving the new
group. After all vertices are merged, the scalar on the left-hand side
of \eqref{eq:graph-contraction-bound} remains, and repeated application
of \eqref{eq:contraction-cauchy-schwarz} proves the claim.
\end{proof}

\subsection{Evaluation of the partition monotones}
\label{subsec:maximal-pin-evaluation}

\begin{theorem}[Maximally entangled PIN evaluation]
\label{thm:maximal-pin-evaluation}
For every partition $\mathcal P=\{B_1,\ldots,B_q\}$,
\begin{equation}
    E_{\mathcal P}(\Psi_G)
    =\frac{\displaystyle\sum_{e\in E(\mathcal P)}\log D_e}{q-1},
    \qquad
    U(\Psi_G)=\tau(G,c),
    \qquad c_e=\log D_e.
    \label{eq:maximal-pin-evaluation}
\end{equation}
\end{theorem}

\begin{proof}
Group the parties according to $\mathcal P$ and discard internal edges,
which supply pure states local to a block. This leaves $E_{\mathcal P}$
unchanged by \eqref{eq:local-invariance}. The crossing edges form a
multigraph with no self loops on the $q$ blocks. Denote its state by
$\Psi_{\mathcal P}=|\Psi_{\mathcal P}\rangle
\langle\Psi_{\mathcal P}|$ and set
\begin{equation}
    P_{\mathcal P}:=\prod_{e\in E(\mathcal P)}D_e,
    \qquad
    r_i:=\prod_{\substack{e\in E(\mathcal P)\\
                          e\cap B_i\neq\varnothing}}D_e.
    \label{eq:crossing-edge-dimensions}
\end{equation}
Here $r_i$ is the local support dimension at block $B_i$. Each
crossing edge contributes to the dimensions of both endpoint blocks, so
\begin{equation}
    \prod_{i=1}^{q}r_i=P_{\mathcal P}^{\,2}.
    \label{eq:pin-support-dimension}
\end{equation}
Work on these local support spaces. For an edge-label assignment
$\mathbf a=(a_e)_{e\in E(\mathcal P)}$, with $0\leq a_e<D_e$, let
$\mathbf a_i$ be its restriction to the crossing edges incident on
$B_i$. Expanding the edge states gives
\begin{equation}
    |\Psi_{\mathcal P}\rangle
    =\frac1{\sqrt{P_{\mathcal P}}}
      \sum_{\mathbf a}\bigotimes_{i=1}^{q}|\mathbf a_i\rangle_i.
    \label{eq:grouped-pin-expansion}
\end{equation}

\emph{Upper bound.}
For a normalized product vector $\bigotimes_i|u_i\rangle$, define
$f_i(\mathbf a_i)=\langle u_i|\mathbf a_i\rangle$. Each array has
$\|f_i\|_2=1$. Lemma~\ref{lem:graph-contraction} therefore gives
\begin{equation}
    \left|\left\langle\bigotimes_i u_i
      \middle|\Psi_{\mathcal P}\right\rangle\right|^2
    =\frac1{P_{\mathcal P}}
      \left|\sum_{\mathbf a}\prod_i f_i(\mathbf a_i)\right|^2
    \leq\frac1{P_{\mathcal P}}.
    \label{eq:pin-product-overlap-bound}
\end{equation}
Choosing every $u_i$ to select the all-zero assignment attains equality,
so
\begin{equation}
    \max_{\|u_i\|=1}
    \left|\left\langle\bigotimes_i u_i
      \middle|\Psi_{\mathcal P}\right\rangle\right|^2
    =P_{\mathcal P}^{-1}.
    \label{eq:pin-product-overlap}
\end{equation}

Spectrally decomposing the local positive factors extends this bound
to every separable positive effect $M$:
\begin{equation}
    \operatorname{Tr}(M\Psi_{\mathcal P})
    \leq\frac{\operatorname{Tr}M}{P_{\mathcal P}}
    =P_{\mathcal P}\operatorname{Tr}\!\left(
      M\frac{I}{P_{\mathcal P}^{\,2}}\right).
    \label{eq:pin-separable-effect-bound}
\end{equation}
Against the product reference $I/P_{\mathcal P}^{\,2}
=\bigotimes_i I_i/r_i$, every nonzero effect of a separable POVM has
likelihood ratio at most $P_{\mathcal P}$. Thus
\begin{equation}
    J_{\mathcal P}(\Psi_{\mathcal P})
    \leq D_{\mathrm{SEP}(\mathcal P)}\!\left(
      \Psi_{\mathcal P}\middle\|\frac{I}{P_{\mathcal P}^{\,2}}
    \right)
    \leq\log P_{\mathcal P}.
    \label{eq:pin-j-upper-bound}
\end{equation}

\emph{Lower bound.}
Measure every crossing-edge register in its computational basis.
Edge $e$ produces an independent uniform label of entropy $\log D_e$,
reported by both endpoint blocks. If $p$ is the joint outcome
distribution and $p_i$ its marginal at block $B_i$, then
\begin{equation}
    H(p)=\sum_{e\in E(\mathcal P)}\log D_e=\log P_{\mathcal P},
    \qquad
    \sum_iH(p_i)=2\log P_{\mathcal P}.
    \label{eq:pin-measurement-entropies}
\end{equation}
The joint distribution counts each label once and the marginals twice.
Lemma~\ref{lem:classical-correlation-bound} gives
\begin{equation}
    J_{\mathcal P}(\Psi_{\mathcal P})\geq\log P_{\mathcal P}.
    \label{eq:pin-j-lower-bound}
\end{equation}
The bounds give $J_{\mathcal P}(\Psi_{\mathcal P})=\log P_{\mathcal P}$.
Purity and \eqref{eq:pure-state-reduction} then give $E_{\mathcal P}$,
and minimizing over partitions gives $U$.
\end{proof}

Since $\Phi_{D_e}^{\otimes n}$ is locally equivalent to $\Phi_{D_e^n}$,
the theorem also gives
\begin{equation}
    U(\Psi_G^{\otimes n})
    =\tau(G,nc)=n\tau(G,c).
    \label{eq:maximal-pin-tensor-powers}
\end{equation}
For arbitrary edge ranks, Corollary~\ref{cor:exact-ghz-converse}
implies that exact conversion to $\Gamma_{V,d}$ requires
\begin{equation}
    \log d\leq\tau(G,c).
    \label{eq:maximal-pin-exact-converse}
\end{equation}
We next establish matching achievability for integer Bell supplies and
qubit GHZ outputs.

\subsection{Spanning-tree packing and GHZ fusion}
\label{subsec:tree-packing-fusion}

A spanning tree connects all vertices without a cycle. The
Nash--Williams--Tutte theorem~\cite{Tutte1961,NashWilliams1961}
states that a multigraph $H$ contains $k$ pairwise edge-disjoint
spanning trees if and only if
\begin{equation}
    |E_H(\mathcal P)|\geq k(|\mathcal P|-1)
    \qquad\text{for every partition }\mathcal P;
    \label{eq:tree-packing-condition}
\end{equation}
Necessity follows because each spanning tree uses at least
$|\mathcal P|-1$ crossing edges; for a short proof of sufficiency,
see Ref.~\cite{Kaiser2012}. We combine this theorem with the following
GHZ-generation protocol from shared Bell pairs~\cite{MeignantEtAl2019}.

\begin{lemma}[Tree fusion]
\label{lem:tree-fusion}
One Bell pair on every edge of a spanning tree can be converted
exactly and deterministically by LOCC into one all-party qubit GHZ
state. Consequently, $k$ edge-disjoint spanning trees yield $k$ GHZ
states.
\end{lemma}

\begin{proof}
Start from one tree edge, whose Bell pair is a GHZ state on its two
endpoints. Suppose a GHZ state already covers a connected vertex set
$S$, and a tree edge joins $v\in S$ to an unvisited vertex $w$.
Write their joint resource as
\[
    |\Gamma_{S,2}\rangle\otimes|\Phi_2\rangle_{b_vb_w}
    =\frac12\sum_{x,y=0}^{1}
      |x\rangle^{\otimes|S|}|y\rangle_{b_v}|y\rangle_{b_w}.
\]
At $v$, apply a CNOT from its GHZ qubit to its Bell register $b_v$,
then measure $b_v$ in the computational basis. Outcome
$z=y\oplus x$ occurs with probability $1/2$ and leaves the normalized
state
\begin{equation}
    \frac1{\sqrt2}\sum_{x=0}^{1}
      |x\rangle^{\otimes|S|}|x\oplus z\rangle_{b_w},
    \label{eq:tree-fusion-branch}
\end{equation}
where $\oplus$ denotes addition modulo two. Communicate $z$ to $w$
and apply the Pauli correction $X^z$. Both outcomes then give
$|\Gamma_{S\cup\{w\},2}\rangle$ exactly. Repeating along the tree
reaches every vertex. Edge-disjoint trees use distinct Bell-pair
registers, so the construction applies separately to each tree.
\end{proof}

\begin{theorem}[Optimal exact Bell-network yield]
\label{thm:optimal-exact-bell-yield}
For integer Bell-pair supplies $m=(m_e)_{e\in E}$,
\begin{equation}
    K_{\mathrm{LOCC}}(G,m)
    =K_{\mathrm{SEP}}(G,m)
    =\lfloor\tau(G,m)\rfloor.
    \label{eq:optimal-exact-bell-yield}
\end{equation}
\end{theorem}

\begin{proof}
For $D_e=2^{m_e}$, Theorem~\ref{thm:maximal-pin-evaluation} gives
$U(\Psi_{G,m})=\tau(G,m)$, so
Corollary~\ref{cor:exact-ghz-converse} yields the converse:
\begin{equation}
    K_{\mathrm{SEP}}(G,m)\leq\lfloor\tau(G,m)\rfloor.
    \label{eq:bell-yield-upper-bound}
\end{equation}

For achievability, replace edge $e$ by $m_e$ individually labelled
parallel edges, each representing one available Bell pair. In this
expanded multigraph $G_m$,
\[
    |E_{G_m}(\mathcal P)|=\sum_{e\in E(\mathcal P)}m_e.
\]
The largest integer satisfying \eqref{eq:tree-packing-condition}
is $k_*:=\lfloor\tau(G,m)\rfloor$. The tree-packing theorem provides
$k_*$ edge-disjoint spanning trees. Fusing them by
Lemma~\ref{lem:tree-fusion} and discarding unused pairs gives a
trace-preserving LOCC protocol with $k_*$ exact GHZ states, hence
\begin{equation}
    K_{\mathrm{LOCC}}(G,m)\geq\lfloor\tau(G,m)\rfloor.
    \label{eq:bell-yield-lower-bound}
\end{equation}
The claim follows from $K_{\mathrm{LOCC}}\leq K_{\mathrm{SEP}}$.
\end{proof}

If the positive-supply graph is disconnected, partitioning it into
its connected components gives $\tau(G,m)=0$, so the formula covers
disconnected networks and vanishing edge supplies.

\subsection{Examples}
\label{subsec:exact-bell-examples}

\paragraph{Trees.}
For a tree $G$ with nonnegative edge weights $c_e$,
\begin{equation}
    \tau(G,c)=\min_{e\in E}c_e.
    \label{eq:tree-partition-strength}
\end{equation}
Indeed, every partition into $q$ blocks has at least $q-1$ crossing
edges, so its ratio is at least $\min_e c_e$. Removing a
least-weight edge gives a bipartition attaining this value. For Bell
supplies, the optimal exact yield is therefore $\min_e m_e$.

\paragraph{Uniform cycles.}
Let $C_N$, $N\geq3$, have $m$ Bell pairs on every edge, and let
$\mathbf1$ denote the all-ones edge vector. Every proper nonempty
vertex set of a cycle has at least two boundary edges. A partition
into $q$ blocks therefore has at least $q$ crossing edges, giving
\[
    \frac{\sum_{e\in E(\mathcal P)}m}{q-1}
    \geq\frac{mq}{q-1}
    \geq\frac{mN}{N-1}.
\]
The singleton partition attains equality, so
\begin{equation}
    \tau(C_N,m\mathbf1)=\frac{Nm}{N-1},
    \qquad
    K_{\mathrm{LOCC}}(C_N,m\mathbf1)
    =\left\lfloor\frac{Nm}{N-1}\right\rfloor.
    \label{eq:cycle-exact-yield}
\end{equation}
For example, four Bell pairs on each pentagon edge yield exactly five
GHZ states. Deleting each cycle edge in turn gives five spanning paths,
with every edge appearing in four paths. Assigning distinct Bell pairs
to these paths and fusing them attains the bound.

\section{PINs with Arbitrary Pure Edges}
\label{sec:independent-pure-edges}

Let each edge supply an arbitrary pure bipartite state,
$\rho_G=\bigotimes_{e\in E}\phi_e$. In local Schmidt bases,
\begin{equation}
    |\phi_e\rangle
    =\sum_{a=1}^{r_e}\sqrt{\lambda_{e,a}}\,
      |a\rangle_{u,e}|a\rangle_{v,e},
    \qquad
    s_e=H(\lambda_e),
    \qquad e=\{u,v\},
    \label{eq:pure-edge-schmidt-form}
\end{equation}
where the positive Schmidt probabilities sum to one. The partition
formula continues to hold with the entanglement entropies $s_e$ as
weights. Its proof uses the product of local marginals as a reference
and an entropy bound for separable positive operators.

\subsection{Evaluation in terms of edge entropies}
\label{subsec:pure-pin-evaluation}

We use operator monotonicity of the logarithm and the operator Jensen
inequality~\cite{HansenPedersen2003}: for a completely positive unital
map $\mathcal A$ (so $\mathcal A(I)=I$),
\begin{equation}
    \mathcal A(\log Y)\leq\log\mathcal A(Y),
    \qquad Y>0.
    \label{eq:operator-log-jensen}
\end{equation}
All inverses and logarithms below are taken on the tensor product of
the local supports of the network state.

\begin{lemma}[Entropy bound for separable positive operators]
\label{lem:separable-operator-entropy-bound}
Let $H=(W,F)$ be a finite multigraph with no self loops and independent pure
bipartite edge states,
\begin{equation}
    \omega_H=\bigotimes_{e\in F}\phi_e,
    \qquad
    \omega_w=\operatorname{Tr}_{W\setminus\{w\}}\omega_H,
    \qquad
    \pi_H=\bigotimes_{w\in W}\omega_w.
    \label{eq:pure-network-marginal-reference}
\end{equation}
For every positive definite operator $X$ that is separable across
$W$ and satisfies $\operatorname{Tr}(\pi_HX)=1$,
\begin{equation}
    \operatorname{Tr}(\omega_H\log X)
    \leq\sum_{e\in F}s_e.
    \label{eq:separable-operator-entropy-bound}
\end{equation}
\end{lemma}

The proof is given in Appendix~\ref{A1}.

\begin{theorem}[Independent pure-edge evaluation]
\label{thm:pure-pin-evaluation}
Let $\rho_G=\bigotimes_{e\in E}\phi_e$ have independent pure
bipartite edges with entanglement entropies $s_e$. For every partition
$\mathcal P=\{B_1,\ldots,B_q\}$,
\begin{equation}
    E_{\mathcal P}(\rho_G)
    =\frac{\displaystyle\sum_{e\in E(\mathcal P)}s_e}{q-1},
    \qquad
    U(\rho_G)=\tau(G,s).
    \label{eq:pure-pin-evaluation}
\end{equation}
For every positive integer $n$,
\begin{equation}
    U(\rho_G^{\otimes n})=n\tau(G,s).
    \label{eq:pure-pin-tensor-powers}
\end{equation}
\end{theorem}

\begin{proof}
Group the parties according to $\mathcal P$ and discard internal pure
edges using \eqref{eq:local-invariance}, as in
Theorem~\ref{thm:maximal-pin-evaluation}. Let $\omega$ be the remaining
network of crossing edges, and set
\begin{equation}
    L_{\mathcal P}:=\sum_{e\in E(\mathcal P)}s_e.
    \label{eq:crossing-edge-entropy}
\end{equation}
Purity and \eqref{eq:pure-state-reduction} give
\begin{equation}
    E_{\mathcal P}(\rho_G)
    =\frac{J_{\mathcal P}(\omega)}{q-1}.
    \label{eq:pure-pin-convexification-reduction}
\end{equation}
We show that $J_{\mathcal P}(\omega)=L_{\mathcal P}$.

\emph{Lower bound.}
Schmidt-basis measurements on the crossing edges produce independent
labels with distributions $\lambda_e$ and entropies $s_e$, each shared
by its endpoint blocks. For the joint distribution $p$ and its block
marginals $p_i$,
\begin{equation}
    H(p)=L_{\mathcal P},
    \qquad
    \sum_iH(p_i)=2L_{\mathcal P}.
    \label{eq:pure-pin-measurement-entropies}
\end{equation}
Lemma~\ref{lem:classical-correlation-bound} therefore yields
\begin{equation}
    J_{\mathcal P}(\omega)\geq L_{\mathcal P}.
    \label{eq:pure-pin-j-lower-bound}
\end{equation}

\emph{Upper bound.}
Let $\omega_i$ be the reduced state at block $B_i$ and choose
\begin{equation}
    \pi:=\bigotimes_{i=1}^{q}\omega_i
    \label{eq:pure-pin-product-reference}
\end{equation}
as the product reference. Work on the local supports, where $\pi>0$.
For an arbitrary separable POVM $\{M_y\}_y$, put
\[
    p_y=\operatorname{Tr}(\omega M_y),
    \qquad
    r_y=\operatorname{Tr}(\pi M_y).
\]
Omit zero effects so that $r_y>0$. The classical variational formula is
\begin{equation}
    D(p\|r)
    =\sup_{\substack{t_y>0\\\sum_y r_yt_y=1}}
      \sum_y p_y\log t_y.
    \label{eq:classical-relative-entropy-variational}
\end{equation}
Writing $\nu_y=r_yt_y$ makes the objective
$D(p\|r)-D(p\|\nu)$; strictly positive distributions $\nu$ can
approach $p$, including when some $p_y$ vanish.

For any admissible $t$, the operator $X:=\sum_y t_yM_y$ is separable,
satisfies $X\geq(\min_y t_y)I>0$, and obeys
$\operatorname{Tr}(\pi X)=\sum_y r_yt_y=1$.
Since $\sum_yM_y=I$, operator Jensen inequality~\cite{HansenPedersen2003} and
Lemma~\ref{lem:separable-operator-entropy-bound} give
\begin{equation}
    \sum_y p_y\log t_y
    \leq\operatorname{Tr}\!\left[
      \omega\log\!\left(\sum_y t_yM_y\right)
    \right]
    \leq L_{\mathcal P}.
    \label{eq:pure-pin-measured-entropy-bound}
\end{equation}
Taking the supremum over $t$ and all separable POVMs gives
\begin{equation}
    J_{\mathcal P}(\omega)
    \leq D_{\mathrm{SEP}(\mathcal P)}(\omega\|\pi)
    \leq L_{\mathcal P}.
    \label{eq:pure-pin-j-upper-bound}
\end{equation}
Together with \eqref{eq:pure-pin-j-lower-bound}, this gives
$J_{\mathcal P}(\omega)=L_{\mathcal P}$. The fixed-partition formula
follows from \eqref{eq:pure-pin-convexification-reduction}, and
minimizing over partitions yields $U(\rho_G)=\tau(G,s)$.

Applying the same evaluation to $\rho_G^{\otimes n}$, whose edge
states $\phi_e^{\otimes n}$ have entropies $ns_e$, gives
\[
    U(\rho_G^{\otimes n})=\tau(G,ns)=n\tau(G,s),
\]
which proves \eqref{eq:pure-pin-tensor-powers}.
\end{proof}

\subsection{Exact deterministic GHZ converse}
\label{subsec:exact-pure-edge-converse}

The evaluation gives the following form of the general converse from
Section~\ref{sec:partition-monotone}.

\begin{corollary}[Exact pure-edge converse]
\label{cor:exact-pure-edge-converse}
For every positive integer $n$,
\begin{equation}
    K_{\mathrm{LOCC}}(\rho_G^{\otimes n})
    \leq K_{\mathrm{SEP}}(\rho_G^{\otimes n})
    \leq\lfloor n\tau(G,s)\rfloor.
    \label{eq:exact-pure-edge-yield-bound}
\end{equation}
\end{corollary}

\begin{proof}
Suppose a physical separable channel converts $\rho_G^{\otimes n}$
exactly into $k_n$ qubit GHZ states. Monotonicity, GHZ normalization,
and Theorem~\ref{thm:pure-pin-evaluation} give
\begin{equation}
    k_n
    =U(\Gamma_{V,2^{k_n}})
    \leq U(\rho_G^{\otimes n})
    =n\tau(G,s).
    \label{eq:exact-pure-edge-converse-chain}
\end{equation}
Since $k_n$ is an integer and $\mathrm{LOCC}\subseteq\mathrm{SEP}$,
the claim follows.
\end{proof}

For general pure edges this is a finite-copy upper bound; the next
section proves matching asymptotic achievability with vanishing error.

\section{Approximate GHZ Conversion}
\label{sec:approximate-ghz-distillation}

To extend the exact converse to an approximate output $\sigma$, we
need a lower bound on $E_{\mathcal P}(\sigma)$ in terms of the error and target
dimension. The dimension dependence matters in asymptotic distillation,
where the output dimension grows with the number of input copies.
All protocols remain trace preserving; approximation concerns the
output state.

\subsection{Entanglement bound for approximate GHZ states}
\label{subsec:near-ghz-estimate}

Write $h_2(t):=-t\log t-(1-t)\log(1-t)$ for the binary entropy,
with $0\log0=0$. The following estimate applies to every ensemble
decomposition and hence controls the mixed convex roof.

\begin{lemma}[Near-target estimate]
\label{lem:near-ghz-estimate}
Consider $q\geq2$ parties, each with a $d$-dimensional register,
where $d\geq2$, and let $\mathcal P$ be their singleton partition.
If $F(\sigma,\Gamma_{q,d})\geq1-\epsilon$, with
$0\leq\epsilon\leq1/4$, then
\begin{equation}
    E_{\mathcal P}(\sigma)
    \geq
    \left(1-\frac{2q}{q-1}\sqrt\epsilon\right)\log d
    -\frac{q+1}{q-1}h_2(\sqrt\epsilon).
    \label{eq:near-ghz-estimate}
\end{equation}
\end{lemma}

\begin{proof}
Take an arbitrary decomposition $\sigma=\sum_\alpha p_\alpha
\sigma_\alpha$ and set
\[
    \epsilon_\alpha
    :=1-\operatorname{Tr}(\Gamma_{q,d}\sigma_\alpha),
    \qquad
    t_\alpha:=T(\sigma_\alpha,\Gamma_{q,d}).
\]
The pure-state trace-distance formula, convexity, and concavity of
the square root give $t_\alpha\leq\sqrt{\epsilon_\alpha}$.
Together with linearity of the overlap, this implies
\begin{equation}
    \bar t:=\sum_\alpha p_\alpha t_\alpha
    \leq\sqrt{\sum_\alpha p_\alpha\epsilon_\alpha}
    =\sqrt{1-F(\sigma,\Gamma_{q,d})}
    \leq\sqrt\epsilon.
    \label{eq:ensemble-average-target-distance}
\end{equation}

Measure each register in the computational basis. Let $P_\alpha$
be the joint distribution on $\sigma_\alpha$, with marginals
$P_{\alpha,i}$, and let $P_*$ be the target distribution. The latter
is uniform on the $d$ common-symbol outcomes $(x,\ldots,x)$, so
\[
    H(P_*)=H(P_{*,i})=\log d.
\]
Measurement and marginalization contract trace distance and total
variation distance, respectively. Thus the joint and marginal
distributions are within total variation distance $t_\alpha$ of
their target distributions.

We use the following consequence of the
classical Fannes--Audenaert inequality~\cite{Audenaert2007}:
for distributions $P,Q$ on an alphabet of size $L\geq2$,
\begin{equation}
    \frac12\|P-Q\|_1\leq t
    \quad\Longrightarrow\quad
    |H(P)-H(Q)|\leq t\log L+h_2(t),
    \qquad 0\leq t\leq1.
    \label{eq:classical-entropy-error-bound}
\end{equation}

The marginal and joint alphabets have sizes $d$ and $d^q$,
respectively, so \eqref{eq:classical-entropy-error-bound} gives
\begin{align*}
    H(P_{\alpha,i})
    &\geq \log d-t_\alpha\log d-h_2(t_\alpha),\\
    H(P_\alpha)
    &\leq \log d+qt_\alpha\log d+h_2(t_\alpha).
\end{align*}

Lemma~\ref{lem:classical-correlation-bound} yields
\begin{align}
    J_{\mathcal P}(\sigma_\alpha)
    &\geq\sum_{i=1}^{q}H(P_{\alpha,i})-H(P_\alpha) \notag\\
    &\geq(q-1)\log d-2qt_\alpha\log d
         -(q+1)h_2(t_\alpha).
    \label{eq:near-ghz-component-bound}
\end{align}
Average over $\alpha$ and use concavity of $h_2$ to obtain
\[
    \sum_\alpha p_\alpha J_{\mathcal P}(\sigma_\alpha)
    \geq(q-1)\log d-2q\bar t\log d-(q+1)h_2(\bar t).
\]
Since $\bar t\leq\sqrt\epsilon\leq1/2$, monotonicity of $h_2$
on $[0,1/2]$ permits replacing $\bar t$ by $\sqrt\epsilon$ on
the right-hand side. Taking the infimum over all decompositions and
dividing by $q-1$ proves \eqref{eq:near-ghz-estimate}.
\end{proof}

\subsection{Finite-error converse for arbitrary inputs}
\label{subsec:finite-error-ghz-converse}

\begin{theorem}[Finite-error GHZ converse]
\label{thm:finite-error-ghz-converse}
Let $\rho$ be an arbitrary joint input state and $\Lambda$ a
trace-preserving channel separable across the physical parties.
If
\begin{equation}
    F\!\left(\Lambda(\rho),\Gamma_{V,2^k}\right)
    \geq1-\epsilon,
    \qquad 0\leq\epsilon\leq1/4,
    \label{eq:finite-error-ghz-assumption}
\end{equation}
then every partition $\mathcal P$ with $q\geq2$ blocks satisfies
\begin{equation}
    E_{\mathcal P}(\rho)
    \geq
    \left(1-\frac{2q}{q-1}\sqrt\epsilon\right)k
    -\frac{q+1}{q-1}h_2(\sqrt\epsilon).
    \label{eq:finite-error-partition-converse}
\end{equation}
Consequently,
\begin{equation}
    U(\rho)\geq(1-4\sqrt\epsilon)k
                  -3h_2(\sqrt\epsilon).
    \label{eq:finite-error-u-converse}
\end{equation}
The same bounds apply, in particular, to LOCC protocols.
\end{theorem}

\begin{proof}
The case $k=0$ is immediate, so take $d=2^k\geq2$. Fix a partition
and group the output parties according to its blocks. At block $B_i$,
compress the subspace spanned by $\{|x\rangle^{\otimes|B_i|}\}_x$
to one $d$-dimensional register, preparing a fixed state on the
orthogonal complement. The resulting block-local channel $\mathcal C$
sends $\Gamma_{V,d}$ to $\Gamma_{q,d}$ and does not decrease the
target overlap:
\begin{equation}
    F\!\left(\mathcal C(\Lambda(\rho)),\Gamma_{q,d}\right)
    \geq F\!\left(\Lambda(\rho),\Gamma_{V,d}\right).
    \label{eq:ghz-compression-fidelity}
\end{equation}
Indeed, the branch projecting every block onto its encoded subspace
contributes exactly the original overlap; all remaining branches
contribute nonnegative amounts.

Since $\Lambda$ and $\mathcal C$ are separable across $\mathcal P$,
monotonicity under these channels and Lemma~\ref{lem:near-ghz-estimate} give
\eqref{eq:finite-error-partition-converse}. Finally,
\[
    \frac{2q}{q-1}\leq4,
    \qquad
    \frac{q+1}{q-1}\leq3.
\]
These estimates make the lower bound uniform over partitions;
minimizing proves \eqref{eq:finite-error-u-converse}.
\end{proof}

For a general joint input, define the
\emph{regularized minimum measured correlation entanglement} by
\[
    U_\infty(\rho)
    :=\liminf_{n\to\infty}
      \frac{1}{n}U(\rho^{\otimes n}).
\]
Using a lower limit avoids assuming additivity or existence of an
ordinary limit. For the independent pure-edge networks and publicly
flagged Bell-network mixtures considered here, the explicit evaluations
give $U(\rho^{\otimes n})=nU(\rho)$ for every positive integer $n$,
so $U_\infty(\rho)=U(\rho)$.
The finite-error estimate gives the asymptotic converse
\begin{equation}
    R_{\mathrm{LOCC}}(\rho)
    \leq R_{\mathrm{SEP}}(\rho)
    \leq U_\infty(\rho).
    \label{eq:general-vanishing-error-converse}
\end{equation}
To verify the second inequality, take an achievable protocol sequence
and write $\epsilon_n$ for its overlap infidelity. Vanishing trace
distance implies $\epsilon_n\to0$. Applying
\eqref{eq:finite-error-u-converse} to $\rho^{\otimes n}$ gives, for
all sufficiently large $n$,
\[
    \frac1nU(\rho^{\otimes n})
    \geq(1-4\sqrt{\epsilon_n})\frac{k_n}{n}
       -\frac3n h_2(\sqrt{\epsilon_n}).
\]
Taking the lower limit and then the supremum over achievable rates
proves \eqref{eq:general-vanishing-error-converse}.

\subsection{Optimal rates for PINs with pure edges}
\label{subsec:optimal-pure-edge-rates}

\begin{theorem}[Pure-edge GHZ distillation capacity]
\label{thm:pure-edge-distillation-capacity}
Let $\rho_G=\bigotimes_{e\in E}\phi_e$ have independent pure
bipartite edges with entanglement entropies $s_e$. Then
\begin{equation}
    R_{\mathrm{LOCC}}(\rho_G)
    =R_{\mathrm{SEP}}(\rho_G)
    =U(\rho_G)=\tau(G,s).
    \label{eq:pure-edge-distillation-capacity}
\end{equation}
\end{theorem}

\begin{proof}
\emph{Converse.}
Theorem~\ref{thm:pure-pin-evaluation} evaluates every tensor power:
$U(\rho_G^{\otimes n})=n\tau(G,s)$. Substitution into
\eqref{eq:general-vanishing-error-converse} gives
\begin{equation}
    R_{\mathrm{SEP}}(\rho_G)\leq\tau(G,s).
    \label{eq:pure-edge-rate-upper-bound}
\end{equation}

\emph{Achievability.}
Fix $\delta>0$ and define deterministic Bell-pair supplies
\begin{equation}
    m_e^{(n)}:=\left\lfloor n(s_e-\delta)_+\right\rfloor,
    \qquad (x)_+:=\max\{x,0\}.
    \label{eq:concentrated-bell-supplies}
\end{equation}
For each edge with $s_e>\delta$, bipartite pure-state entanglement
concentration produces $m_e^{(n)}$ exact Bell pairs on a heralded
success event whose failure probability $\beta_{e,n}$ tends to
zero~\cite{BennettEtAl1996}. Edges with $s_e\leq\delta$ require no
pairs and do not fail. Concentration on the distinct edge registers
is a physical LOCC protocol. Finiteness of the edge set gives
the following bound on the probability $\beta_n$ that any edge fails:
\begin{equation}
    \beta_n\leq\sum_{e\in E}\beta_{e,n}\longrightarrow0.
    \label{eq:network-concentration-failure}
\end{equation}

When every edge succeeds, retain the prescribed supplies and apply
the tree-packing and fusion protocol of
Theorem~\ref{thm:optimal-exact-bell-yield}. It produces exactly
\begin{equation}
    k_n:=\left\lfloor\tau(G,m^{(n)})\right\rfloor
    \label{eq:concentration-fusion-output}
\end{equation}
qubit GHZ states. On failure, prepare a fixed product state on the
same output registers. The complete protocol is trace preserving,
and its output $\sigma_n$ after discarding the success flag obeys
\begin{equation}
    T(\sigma_n,\Gamma_{V,2^{k_n}})\leq\beta_n\longrightarrow0.
    \label{eq:concentration-fusion-error}
\end{equation}

By homogeneity of $\tau$ and its continuity as the minimum of
finitely many linear functions,
\begin{equation}
    \lim_{n\to\infty}\frac{k_n}{n}
    =\tau\!\left(G,(s-\delta)_+\right).
    \label{eq:concentration-fusion-rate}
\end{equation}
Letting $\delta\downarrow0$ proves
$R_{\mathrm{LOCC}}(\rho_G)\geq\tau(G,s)$.
Together with \eqref{eq:pure-edge-rate-upper-bound} and
Theorem~\ref{thm:pure-pin-evaluation}, this proves
\eqref{eq:pure-edge-distillation-capacity}.
\end{proof}

This establishes \eqref{eq:main-pure-capacity}. In particular,
maximally entangled edges have $s_e=\log D_e$, so
\begin{equation}
    R_{\mathrm{LOCC}}(\Psi_G)
    =R_{\mathrm{SEP}}(\Psi_G)
    =\tau\!\left(G,\{\log D_e\}_{e\in E}\right).
    \label{eq:maximal-pin-distillation-capacity}
\end{equation}

\section{Locally Flagged Correlated Mixed Networks}
\label{sec:flagged-correlated-networks}

A shared classical label can correlate the edge resources. For the
family below, the partition monotones and optimal GHZ rate remain
exactly evaluable: pooling Bell pairs across copies combines
complementary connections from different configurations.

\subsection{Flagged resources and optimal rates}
\label{subsec:flagged-resources-rates}

Let $x$ range over a finite set of configurations with probabilities
$p_x$. Configuration $x$ supplies $m_e^{(x)}\in\mathbb N_0$ Bell
pairs on edge $e$ of a fixed multigraph $G=(V,E)$. Embed all
configurations in common local Hilbert spaces by padding unused
registers with fixed local pure states. Give each party an orthogonal
classical flag recording $x$, and define
\begin{equation}
    \rho_{\mathrm{flag}}
    :=\sum_xp_x
      \left(\bigotimes_{v\in V}|x\rangle\langle x|_{X_v}\right)
      \otimes\Psi_{G,m^{(x)}},
    \qquad
    \bar m_e:=\sum_xp_xm_e^{(x)}.
    \label{eq:publicly-flagged-network}
\end{equation}
Every party can read the label without disturbing the conditional Bell
pairs. The edge supplies may be arbitrarily correlated through $x$,
so the input need not factor into independent edge states.

\begin{theorem}[Locally flagged correlated Bell networks]
\label{thm:flagged-network-capacity}
For every partition $\mathcal P$ with $q\geq2$ blocks and every
positive integer $n$,
\begin{equation}
    E_{\mathcal P}(\rho_{\mathrm{flag}}^{\otimes n})
    =n\,\frac{\displaystyle\sum_{e\in E(\mathcal P)}\bar m_e}{q-1},
    \qquad
    U(\rho_{\mathrm{flag}}^{\otimes n})=n\tau(G,\bar m).
    \label{eq:flagged-network-evaluation}
\end{equation}
Moreover,
\begin{equation}
    R_{\mathrm{LOCC}}(\rho_{\mathrm{flag}})
    =R_{\mathrm{SEP}}(\rho_{\mathrm{flag}})
    =U_\infty(\rho_{\mathrm{flag}})
    =U(\rho_{\mathrm{flag}})=\tau(G,\bar m).
    \label{eq:flagged-network-capacity}
\end{equation}
\end{theorem}

\begin{proof}
\emph{Evaluation and converse.}
Reading a flag produces configuration $x$ with probability $p_x$.
Selective monotonicity and local ancilla invariance give
$E_{\mathcal P}(\rho_{\mathrm{flag}})
 \geq\sum_xp_xE_{\mathcal P}(\Psi_{G,m^{(x)}})$.
Convexity gives the reverse inequality because each fixed flag is a
product of local pure states. Thus the fixed-partition monotones obey
the FLAGS identity for locally orthogonal classical
labels~\cite{Horodecki2005}:
\begin{equation}
    E_{\mathcal P}(\rho_{\mathrm{flag}})
    =\sum_xp_xE_{\mathcal P}(\Psi_{G,m^{(x)}})
    =\frac{\displaystyle\sum_{e\in E(\mathcal P)}\bar m_e}{q-1},
    \label{eq:flagged-partition-average}
\end{equation}
where the last equality uses
Theorem~\ref{thm:maximal-pin-evaluation} and local padding invariance.

For $n$ copies, a flag sequence $x^n=(x_1,\ldots,x_n)$ has
probability $\prod_{j=1}^{n}p_{x_j}$. Up to local register
permutations, its conditional resource is a Bell network with
pooled supplies
\begin{equation}
    M_e(x^n):=\sum_{j=1}^{n}m_e^{(x_j)}.
    \label{eq:pooled-flagged-supplies}
\end{equation}
The FLAGS identity for the sequence, with mean supplies $n\bar m_e$,
gives the first identity in \eqref{eq:flagged-network-evaluation}.
Minimizing over partitions gives the second. Hence
$U_\infty(\rho_{\mathrm{flag}})=U(\rho_{\mathrm{flag}})
=\tau(G,\bar m)$, and
\eqref{eq:general-vanishing-error-converse} gives
$R_{\mathrm{SEP}}(\rho_{\mathrm{flag}})\leq\tau(G,\bar m)$.

\emph{Achievability by pooling.}
Read the flags of all input copies and collect the Bell pairs
available on each edge. Fix $\eta>0$ and set
\begin{equation}
    \ell_e^{(n)}:=\left\lfloor n(\bar m_e-\eta)_+\right\rfloor,
    \qquad (a)_+:=\max\{a,0\}.
    \label{eq:flagged-target-supplies}
\end{equation}
Independence of the input copies, the law of large numbers, and
finiteness of $E$ imply
\begin{equation}
    \beta_n:=\Pr\!\left\{
      M_e(x^n)<\ell_e^{(n)}\text{ for some }e\in E
    \right\}\longrightarrow0.
    \label{eq:flagged-pooling-failure}
\end{equation}
On the complementary event, retain $\ell_e^{(n)}$ pairs per edge and apply
Theorem~\ref{thm:optimal-exact-bell-yield} to produce
$k_n:=\lfloor\tau(G,\ell^{(n)})\rfloor$ GHZ states. On failure,
discard the resources and prepare a fixed product state on the same
output registers. The resulting trace-preserving physical LOCC
protocol has output $\sigma_n$ satisfying
\begin{equation}
    T(\sigma_n,\Gamma_{V,2^{k_n}})\leq\beta_n\longrightarrow0.
    \label{eq:flagged-pooling-output-error}
\end{equation}
Homogeneity and continuity of $\tau$ give
\[
    \lim_{n\to\infty}\frac{k_n}{n}
    =\tau\!\left(G,(\bar m-\eta)_+\right).
\]
Letting $\eta\downarrow0$ proves
$R_{\mathrm{LOCC}}(\rho_{\mathrm{flag}})\geq\tau(G,\bar m)$.
Together with the converse and
$R_{\mathrm{LOCC}}\leq R_{\mathrm{SEP}}$, this proves
\eqref{eq:flagged-network-capacity}.
\end{proof}

\subsection{Pooling and activation}
\label{subsec:flagged-network-activation}

Consider three parties $A,B,C$. With probability $1/2$, the
configuration supplies one Bell pair on $AB$ and leaves $C$
unentangled; otherwise it supplies one Bell pair on $BC$ and leaves
$A$ unentangled. The local flag identifies the configuration.
Every conditional state is separable across a nontrivial cut and
therefore has zero all-party GHZ distillation rate. Nevertheless,
\begin{equation}
    \bar m_{AB}=\bar m_{BC}=\frac12,
    \qquad
    R_{\mathrm{LOCC}}(\rho_{\mathrm{flag}})
    =R_{\mathrm{SEP}}(\rho_{\mathrm{flag}})
    =U(\rho_{\mathrm{flag}})=\frac12.
    \label{eq:flagged-path-activation}
\end{equation}
Indeed, a sequence containing $n_{AB}$ configurations of the first
type and $n_{BC}$ of the second supplies a path with those Bell-pair
counts. Pairing an $AB$ resource with a $BC$ resource and fusing
at $B$ gives $\min\{n_{AB},n_{BC}\}$ GHZ states. Both counts are
asymptotically $n/2$.

A protocol processing each copy separately cannot generate a GHZ
state from either configuration. Pooling complementary connections
across copies connects all three parties; the flags identify the
resources but carry no entanglement.

The flagged state is biseparable: its conditional states are separable
across $AB|C$ and $A|BC$, respectively. Its positive asymptotic GHZ
rate illustrates the tensor instability of biseparability discussed
by Murray et al.~\cite{MurrayEtAl2026}: biseparability need not be
preserved under tensor powers when each party groups its copies.
The mixture is entangled across every fixed bipartition, so
$U(\rho_{\mathrm{flag}})=1/2$ agrees with
Proposition~\ref{prop:faithfulness}.

\subsection{Averaging before partition minimization}
\label{subsec:flagged-partition-order}

The FLAGS identity applies at each fixed partition. Consequently,
\begin{align}
    U(\rho_{\mathrm{flag}})
    &=\min_{\mathcal P}\sum_xp_x
      E_{\mathcal P}(\Psi_{G,m^{(x)}}) \notag\\
    &\geq\sum_xp_x\min_{\mathcal P}
      E_{\mathcal P}(\Psi_{G,m^{(x)}})
      =\sum_xp_xU(\Psi_{G,m^{(x)}}).
    \label{eq:flagged-minimum-average-order}
\end{align}
The inequality is strict in the preceding example, where every
conditional network is disconnected:
\begin{equation}
    \min_{\mathcal P}\sum_xp_x
      E_{\mathcal P}(\Psi_{G,m^{(x)}})=\frac12,
    \qquad
    \sum_xp_x\min_{\mathcal P}
      E_{\mathcal P}(\Psi_{G,m^{(x)}})=0.
    \label{eq:flagged-strict-minimum-average-gap}
\end{equation}
Different configurations may have different minimizing partitions.
Averaging at a fixed partition before minimizing retains the
complementary connections used in pooling; minimizing each
configuration separately loses this information.

Thus $U$ need not be convex or satisfy the FLAGS equality, although
every $E_{\mathcal P}$ does. Since fixed flags are local product
ancillas, the example mixes states with $U=0$ into a state with
$U=1/2$. Convexifying $U$ would erase this positive one-copy value.
The definition therefore convexifies at each fixed partition before
taking the partition minimum.

\section{Strong Converses}
\label{sec:instrument-strong-converses}

We now strengthen the capacity result to a strong converse: the target
fidelity vanishes at every fixed rate above capacity. For maximally
entangled PINs, applying a complete separable instrument to the resource
and a product comparison state gives two normalized branch distributions.
Combining these with a branchwise amplitude bound yields explicit
exponential decay. Entanglement dilution extends the threshold to
arbitrary independent pure edges.

\subsection{Fidelity bound from instrument completeness}
\label{subsec:normalized-instrument-bound}

Let $|\psi\rangle$ be a normalized pure state shared by $q\geq2$
parties. For party $i$, let $H_i$ be the support of its
reduced state and let $r_i=\dim H_i$. Define
\begin{equation}
    \Delta:=\prod_{i=1}^{q}r_i,
    \qquad
    \lambda(\psi):=
      \max_{\substack{|u_i\rangle\in H_i\\\|u_i\|=1}}
      \left|\left\langle\bigotimes_{i=1}^{q}u_i
        \middle|\psi\right\rangle\right|^2.
    \label{eq:instrument-support-and-overlap}
\end{equation}
Thus $\Delta$ is the dimension of the product of local supports,
and $\lambda(\psi)$ is the largest squared product overlap.

\begin{lemma}[Normalized product-overlap bound]
\label{lem:normalized-product-overlap}
For every trace-preserving channel $\Lambda$ that is separable across
these parties and has one $d$-dimensional output register per party,
with $d\geq2$,
\begin{equation}
    F\!\left(\Lambda(|\psi\rangle\langle\psi|),\Gamma_{q,d}\right)
    \leq\min\left\{
      1,\frac{[\Delta\lambda(\psi)]^{1/(q-1)}}{d}
    \right\}.
    \label{eq:normalized-product-overlap-bound}
\end{equation}
\end{lemma}

\begin{proof}
Restrict each local input to $H_i$; the product Kraus operators
remain complete on $\bigotimes_iH_i$. On this space choose the
normalized product comparison state
\begin{equation}
    \Omega:=\bigotimes_{i=1}^{q}\frac{I_{H_i}}{r_i}
           =\frac{I}{\Delta}.
    \label{eq:instrument-comparison-state}
\end{equation}
Refine the channel into product Kraus operators
$K_z=\bigotimes_iK_{i,z}$ and define
\begin{align}
    p_z&:=\|K_z|\psi\rangle\|^2, \notag\\
    r_z&:=\operatorname{Tr}(K_z\Omega K_z^\dagger)
         =\frac1\Delta\prod_{i=1}^{q}
           \operatorname{Tr}(K_{i,z}^\dagger K_{i,z}), \notag\\
    f_z&:=|\langle\Gamma_{q,d}|K_z|\psi\rangle|^2.
    \label{eq:instrument-branch-data}
\end{align}
For $p_z>0$, the conditional target fidelity is $f_z/p_z$.
Completeness gives
\begin{equation}
    \sum_zp_z=\sum_zr_z=1,
    \qquad F=\sum_zf_z.
    \label{eq:two-input-normalization}
\end{equation}

Fix a branch and write
\[
    b_x:=\langle x|^{\otimes q}K_z|\psi\rangle,
    \qquad
    h_i(x):=\|\langle x|K_{i,z}\|^2,
    \qquad 0\leq x<d.
\]
Normalizing each nonzero local row and using the definition of
$\lambda(\psi)$ gives
\begin{equation}
    |b_x|^2\leq\lambda(\psi)\prod_{i=1}^{q}h_i(x),
    \qquad
    \sum_xh_i(x)=\operatorname{Tr}(K_{i,z}^\dagger K_{i,z}).
    \label{eq:branch-product-row-bound}
\end{equation}
The same inequality holds if a row vanishes, since then $b_x=0$.
H\"older's inequality over $x$ therefore yields
\begin{equation}
    \left(\sum_x|b_x|^{2/q}\right)^q
    \leq\lambda(\psi)\prod_{i=1}^{q}\sum_xh_i(x)
    =\Delta\lambda(\psi)r_z.
    \label{eq:branch-amplitude-comparison-bound}
\end{equation}
The branch probability and target overlap also satisfy
\begin{equation}
    \sum_x|b_x|^2\leq p_z,
    \qquad
    df_z=\left|\sum_xb_x\right|^2
         \leq\left(\sum_x|b_x|\right)^2.
    \label{eq:branch-probability-and-overlap}
\end{equation}

For $q>2$, H\"older interpolation gives
\begin{equation}
    \left(\sum_xa_x\right)^{2q-2}
    \leq\left(\sum_xa_x^{2/q}\right)^q
        \left(\sum_xa_x^2\right)^{q-2},
    \qquad a_x\geq0.
    \label{eq:ghz-amplitude-interpolation}
\end{equation}
Indeed, factor
$a_x=(a_x^{2/q})^{q/(2q-2)}
     (a_x^2)^{(q-2)/(2q-2)}$
and apply H\"older with conjugate exponents
$(2q-2)/q$ and $(2q-2)/(q-2)$.
Combining \eqref{eq:branch-amplitude-comparison-bound}--
\eqref{eq:ghz-amplitude-interpolation} gives
\begin{equation}
    d^{q-1}f_z^{q-1}
    \leq\Delta\lambda(\psi)r_zp_z^{q-2}.
    \label{eq:branchwise-ghz-fidelity-bound}
\end{equation}
Taking $(q-1)$st roots and summing over branches, we obtain by
H\"older and \eqref{eq:two-input-normalization}
\begin{align}
    F
    &\leq\frac{[\Delta\lambda(\psi)]^{1/(q-1)}}{d}
      \sum_zp_z^{(q-2)/(q-1)}r_z^{1/(q-1)} \notag\\
    &\leq\frac{[\Delta\lambda(\psi)]^{1/(q-1)}}{d}
      \left(\sum_zp_z\right)^{(q-2)/(q-1)}
      \left(\sum_zr_z\right)^{1/(q-1)} \notag\\
    &=\frac{[\Delta\lambda(\psi)]^{1/(q-1)}}{d}.
    \label{eq:complete-instrument-fidelity-bound}
\end{align}
For $q=2$, \eqref{eq:branch-amplitude-comparison-bound} and
\eqref{eq:branch-probability-and-overlap} directly give
$df_z\leq\Delta\lambda(\psi)r_z$; summing proves the same bound.
No division by a branch probability is required, so zero-probability
branches are included. Finally, $F\leq1$ completes the proof.
\end{proof}

\subsection{Strong converse for maximally entangled PINs}
\label{subsec:maximal-pin-strong-converse}

\begin{theorem}[Maximally entangled PIN fidelity bound]
\label{thm:maximal-pin-fidelity-bound}
Let $\Psi_G$ have independent maximally entangled edges of ranks
$D_e$, and put $c_e=\log D_e$. Every trace-preserving channel
$\Lambda$ separable across the physical parties satisfies
\begin{equation}
    F\!\left(\Lambda(\Psi_G),\Gamma_{V,d}\right)
    \leq\min\left\{1,\frac{2^{\tau(G,c)}}{d}\right\}.
    \label{eq:maximal-pin-fidelity-bound}
\end{equation}
\end{theorem}

\begin{proof}
Fix $\mathcal P=\{B_1,\ldots,B_q\}$. Since internal pure edges can
be prepared within each block, the original protocol can be simulated
from $\Psi_{\mathcal P}$ by a channel separable across the blocks. Set
\[
    P_{\mathcal P}:=\prod_{e\in E(\mathcal P)}D_e.
\]
Equations~\eqref{eq:pin-product-overlap} and
\eqref{eq:pin-support-dimension} give
\begin{equation}
    \lambda(\Psi_{\mathcal P})=P_{\mathcal P}^{-1},
    \qquad
    \Delta_{\mathcal P}=P_{\mathcal P}^{\,2}.
    \label{eq:grouped-pin-overlap-dimension}
\end{equation}
For every $d$, the block-local compression from the proof of
Theorem~\ref{thm:finite-error-ghz-converse} gives one $d$-dimensional
output register per block without decreasing target overlap, as in
\eqref{eq:ghz-compression-fidelity}. Thus
Lemma~\ref{lem:normalized-product-overlap} yields
\begin{equation}
    F\!\left(\Lambda(\Psi_G),\Gamma_{V,d}\right)
    \leq\frac{P_{\mathcal P}^{1/(q-1)}}{d}
    =\frac1d\,
      2^{\frac{1}{q-1}\sum_{e\in E(\mathcal P)}c_e}.
    \label{eq:partition-instrument-fidelity-bound}
\end{equation}
Minimizing over partitions and using $F\leq1$ proves the theorem.
\end{proof}

For an exact output with $d=2^k$, \eqref{eq:maximal-pin-fidelity-bound}
recovers $k\leq\tau(G,c)$. On $n$ input copies, the edge ranks are
$D_e^n$, and homogeneity of $\tau$ gives
\begin{equation}
    F_n:=F\!\left(\Lambda_n(\Psi_G^{\otimes n}),
                  \Gamma_{V,2^{k_n}}\right)
    \leq\min\{1,2^{n\tau(G,c)-k_n}\}.
    \label{eq:maximal-pin-exponential-fidelity-bound}
\end{equation}
Consequently, if $k_n/n\geq\tau(G,c)+\delta$ for all sufficiently
large $n$ and some fixed $\delta>0$, then
\begin{equation}
    F_n\leq2^{-n\delta}\longrightarrow0,
    \qquad
    T\!\left(\Lambda_n(\Psi_G^{\otimes n}),
             \Gamma_{V,2^{k_n}}\right)
    \geq1-F_n\longrightarrow1.
    \label{eq:maximal-pin-exponential-strong-converse}
\end{equation}
Together with achievability from
Section~\ref{sec:approximate-ghz-distillation}, this identifies
$\tau(G,c)$ as the strong-converse threshold.

\subsection{Strong converse for PINs with arbitrary pure edges}
\label{subsec:pure-edge-strong-converse}

We transfer the threshold to arbitrary pure edges by preparing the
input approximately from a Bell network before applying the conversion
channel.

\begin{theorem}[Pure-edge strong converse]
\label{thm:pure-edge-strong-converse}
Let $\rho_G=\bigotimes_{e\in E}\phi_e$ have independent pure
bipartite edges with entanglement entropies $s_e$. For every sequence
of trace-preserving physical separable channels $\Lambda_n$, if
\begin{equation}
    \frac{k_n}{n}\geq\tau(G,s)+\delta
    \label{eq:pure-edge-above-threshold-rate}
\end{equation}
for all sufficiently large $n$ and some fixed $\delta>0$, then
\begin{equation}
    F\!\left(\Lambda_n(\rho_G^{\otimes n}),
             \Gamma_{V,2^{k_n}}\right)\longrightarrow0,
    \qquad
    T\!\left(\Lambda_n(\rho_G^{\otimes n}),
             \Gamma_{V,2^{k_n}}\right)\longrightarrow1.
    \label{eq:pure-edge-strong-converse}
\end{equation}
\end{theorem}

\begin{proof}
Write $M=|E|$ and choose $\eta>0$ such that $M\eta<\delta/2$.
Entanglement dilution on each edge uses Bell supplies
\begin{equation}
    m_e^{(n)}:=\lceil n(s_e+\eta)\rceil
    \label{eq:dilution-bell-supplies}
\end{equation}
to prepare a network state $\widetilde\rho_n$ satisfying
\begin{equation}
    \epsilon_n:=T(\widetilde\rho_n,\rho_G^{\otimes n})
    \longrightarrow0.
    \label{eq:dilution-network-error}
\end{equation}
In the standard pure-state dilution construction~\cite{BennettEtAl1996},
one endpoint truncates the $n$-copy Schmidt expansion to
a typical set of at most $2^{n(s_e+\eta)}$ labels, prepares the
normalized truncated state locally, and teleports the other half
using $m_e^{(n)}$ Bell pairs. The retained probability tends to one,
so each edge error vanishes; their finite sum bounds the network
error. This preparation defines a trace-preserving physical LOCC
channel $\mathcal D_n$.

The composition $\Lambda_n\circ\mathcal D_n$ is a physical separable
channel from the Bell network $\Psi_{G,m^{(n)}}$. By
Theorem~\ref{thm:maximal-pin-fidelity-bound},
\[
    F\!\left(\Lambda_n(\widetilde\rho_n),
             \Gamma_{V,2^{k_n}}\right)
    \leq 2^{\tau(G,m^{(n)})-k_n}.
\]
Trace-distance contractivity and
$|\operatorname{Tr}[\Gamma(\sigma-\omega)]|\leq T(\sigma,\omega)$
transfer this bound to the true input:
\begin{equation}
    F_n:=F\!\left(\Lambda_n(\rho_G^{\otimes n}),
                  \Gamma_{V,2^{k_n}}\right)
    \leq2^{\tau(G,m^{(n)})-k_n}+\epsilon_n.
    \label{eq:dilution-transferred-fidelity-bound}
\end{equation}

To bound the preparation overhead, choose a partition $\mathcal P_*$
attaining $\tau(G,s)$ and write $q_*=|\mathcal P_*|$. Since
$m_e^{(n)}\leq n(s_e+\eta)+1$,
\begin{align}
    \tau(G,m^{(n)})
    &\leq\frac{\sum_{e\in E(\mathcal P_*)}m_e^{(n)}}{q_*-1}
      \notag\\
    &\leq n\tau(G,s)
      +(n\eta+1)\frac{|E(\mathcal P_*)|}{q_*-1} \notag\\
    &\leq n\tau(G,s)+nM\eta+M.
    \label{eq:dilution-partition-overhead}
\end{align}
Combining this estimate with
\eqref{eq:pure-edge-above-threshold-rate} and
\eqref{eq:dilution-transferred-fidelity-bound} gives, for all
sufficiently large $n$,
\begin{equation}
    F_n\leq2^{M-n\delta/2}+\epsilon_n\longrightarrow0.
    \label{eq:pure-edge-fidelity-decay}
\end{equation}
Finally, $T\geq1-F_n$ proves the trace-distance conclusion.
\end{proof}

Together with Theorem~\ref{thm:pure-edge-distillation-capacity}, this
establishes $\tau(G,s)$ as the strong-converse threshold under LOCC
and separable operations. Equation~\eqref{eq:maximal-pin-exponential-fidelity-bound}
gives explicit exponential decay for maximally entangled PINs;
for arbitrary pure edges, \eqref{eq:pure-edge-fidelity-decay} includes
an additional vanishing dilution error.

\section{Conclusion}
We have introduced the minimum measured correlation entanglement $U(\rho)$, a partition-based entanglement monotone that provides
converses for exact and approximate GHZ distillation from arbitrary multipartite inputs under separable operations, and hence under LOCC. The resulting converses constrain exact
deterministic conversion, approximate preparation at finite error, and asymptotic distillation with vanishing error. The exact yield is bounded by $U(\rho)$, while the asymptotic rate is bounded by its regularized form $U_\infty(\rho)$.

The evaluations give optimal asymptotic rates for independent pure-edge
networks and locally flagged correlated Bell networks, and optimal exact
deterministic yields for fixed integer Bell-pair supplies. A complementary fidelity argument establishes a strong converse
for independent pure-edge networks, with explicit exponential decay for maximally entangled networks. These results bound the entanglement supply in conference-key and quantum secret-sharing protocols that first distill GHZ states~\cite{EppingEtAl2017,HilleryEtAl1999}. They also give lower bounds on the resources needed to prepare GHZ probes for distributed sensing, including entangled clock networks~\cite{KomarEtAl2014}.

\section*{Acknowledgements}

We thank Koji Azuma, Stefan B{\"a}uml, Wojciech Roga, and Masahiro Takeoka
for helpful discussions. We also thank Wojciech Roga and
Masahiro Takeoka for bringing Hua Sun's paper~\cite{Sun2026}
(submitted on 24 November 2026) to our attention and for sharing
a draft of their work on an alternative upper bound for GHZ
distillation~\cite{ishihara2026fundamental}. We will include a
comparison with the latter work in the second version of this
manuscript, once their preprint is publicly available.

The main argument  was obtained with the help of ChatGPT 6.1 (OpenAI). We take full responsibility for the validity of the arguments.

A part of this work was supported by Council for Science,
Technology and Innovation (CSTI), Cross-ministerial Strategic
Innovation Promotion Program (SIP), ``Promoting Application
of Advanced Quantum Technologies to Social Challenges''
(Project management agency: QST; Grant No.~JPJ012367).

\printbibliography

\newpage

\appendix

\section{Proof of Lemma 5.1}
\label{A1}

\begin{proof}
We argue by induction on $|W|$. If there are no edges, the local supports are
one-dimensional and normalization forces $X=1$, so both sides of
\eqref{eq:separable-operator-entropy-bound} vanish.

Fix $v\in W$. For $A_v\geq0$, positivity gives
\[
    \omega_v^{1/2}A_v\omega_v^{1/2}
    \leq\operatorname{Tr}(\omega_vA_v)I_v.
\]
Since $\omega_v$ is invertible on its support, conjugating by
$\omega_v^{-1/2}$ gives
\begin{equation}
    A_v\leq\operatorname{Tr}(\omega_vA_v)\,\omega_v^{-1}.
    \label{eq:weighted-local-operator-bound}
\end{equation}
With
\begin{equation}
    R:=\operatorname{Tr}_v\!\left[
      (\omega_v^{1/2}\otimes I)X
      (\omega_v^{1/2}\otimes I)
    \right],
    \label{eq:weighted-partial-trace-r}
\end{equation}
termwise application of this inequality to a separable decomposition
of $X$ gives $X\leq\omega_v^{-1}\otimes R$. The operator $R$ is
separable across the remaining parties: its
product terms have nonnegative coefficients $\operatorname{Tr}(\omega_vA_v)$.
Choose $\mu>0$ with $X\geq\mu I$. Then $R\geq\mu I$ because
$\operatorname{Tr}\omega_v=1$.

Operator monotonicity gives
\[
    \log X
    \leq -\log\omega_v\otimes I+I_v\otimes\log R.
\]
Taking the expectation in $\omega_H$ yields
\begin{equation}
    \operatorname{Tr}(\omega_H\log X)
    \leq S(\omega_v)
       +\operatorname{Tr}\!\left[
          (\operatorname{Tr}_v\omega_H)\log R
        \right].
    \label{eq:entropy-bound-vertex-removal}
\end{equation}

Tracing out $v$ leaves local mixed states at the other endpoints of
its incident edges. Let $B$ collect these registers and $C$ the
registers of the remaining network $H-v$. Edge independence gives,
up to local register reorderings,
\begin{equation}
    \operatorname{Tr}_v\omega_H=\chi_B\otimes\omega_{H-v},
    \qquad
    \bigotimes_{w\neq v}\omega_w=\chi_B\otimes\pi_{H-v}.
    \label{eq:removed-vertex-factorization}
\end{equation}
Here $\chi_B=\bigotimes_{w\neq v}\chi_{B_w}$ is product across the
remaining parties.

Define the map on operators
\begin{equation}
    \mathcal A_\chi(Y)
    :=\operatorname{Tr}_B\!\left[
      (\chi_B^{1/2}\otimes I_C)Y
      (\chi_B^{1/2}\otimes I_C)
    \right],
    \qquad
    \widetilde R:=\mathcal A_\chi(R).
    \label{eq:local-averaging-map}
\end{equation}
Because $\chi_B$ is product, $\mathcal A_\chi$ is a product of local
completely positive unital maps. Thus $\widetilde R$ is separable
across $H-v$ and satisfies $\widetilde R\geq\mu I>0$. Its normalization is
\begin{align}
    \operatorname{Tr}(\pi_{H-v}\widetilde R)
    &=\operatorname{Tr}\!\left[
        (\chi_B\otimes\pi_{H-v})R
      \right] \notag\\
    &=\operatorname{Tr}(\pi_HX)=1.
    \label{eq:reduced-operator-normalization}
\end{align}
Jensen's inequality~\eqref{eq:operator-log-jensen} and the induction
hypothesis applied to $\widetilde R$ therefore give
\begin{align}
    \operatorname{Tr}\!\left[
      (\operatorname{Tr}_v\omega_H)\log R
    \right]
    &=\operatorname{Tr}\!\left[
        \omega_{H-v}\mathcal A_\chi(\log R)
      \right] \notag\\
    &\leq\operatorname{Tr}(\omega_{H-v}\log\widetilde R)
      \notag\\
    &\leq\sum_{\substack{e\in F\\v\notin e}}s_e.
    \label{eq:entropy-bound-induction}
\end{align}
Since $\omega_v$ is a tensor product of the reduced states of its
incident pure edges,
\[
    S(\omega_v)=\sum_{\substack{e\in F\\v\in e}}s_e.
\]
Substitution into \eqref{eq:entropy-bound-vertex-removal} counts every
edge once and completes the induction.
\end{proof}
\end{document}